\documentclass[preprint,authoryear,12pt]{elsarticle}
\pdfoutput=1

\usepackage{graphicx}

\usepackage{amssymb}

\journal{New Astronomy Reviews}

\begin{document}

\begin{frontmatter}



\title{Galactic Archaeology. The dwarfs that survived and perished}


\author[1]{Vasily Belokurov\corref{cor1}}
\ead{vasily@ast.cam.ac.uk}
\cortext[cor1]{Corresponding author}
\address[1]{Institute of Astronomy, Cambridge}

\begin{abstract} From the archaeological point of view, the local dwarf
galaxies are unique objects in which the imprint of the conditions
that shaped the early structure formation can be studied today at high
resolution. Over the last decade, this new window into the high
redshift Universe has started to be exploited using deep wide-field
imaging, high resolution spectroscopy and cutting edge N-body and
hydro-dynamical simulations. We review the recent advances in the
observational studies of the Milky Way dwarf galaxies, with the aim to
understand the properties of the population as a whole and to assist
an objective comparison between the models and the data.

\end{abstract}


\end{frontmatter}

\let\clearpage\relax

\section{Introduction}
\label{sec:intro}

The prehistoric stars whose formation epochs lie beyond the redshift
accessible by the Hubble Ultra Deep Field, have been found en masse in
little satellites around the Milky Way. Other less fortunate dwarf
galaxies have been pulled apart by gravity to furnish the diffuse
Galactic halo. These recently uncovered relics of the ancient dwarf
galaxy population may play a vital role in the pursuit of
reconstructing the formation of the Galaxy. Its path from the distant
pre-reionisation era, through the most active growth periods to the
present day, can be gleaned by studying the chemical composition and
the phase-space density distribution of these halo denizens. How this
Galactic {\it archaeological} record is collected and analyzed, and
what aspects of galaxy (dwarf and otherwise) formation and evolution
it illuminates is the topic of this review.

The field of Galactic and Local Group studies has enjoyed a decade of
unprecedented busyness thanks to the abundance of data supplied by
several generations of wide-angle sky surveys and the coming of age of
N-body and hydro-dynamic computer simulations of galaxy
formation. Accordingly, there have been several fresh in-depth reviews
of the matters relevant to the topic of this article. In particular,
the star-formation histories and abundances of the Milky Way dwarf
galaxies have been scrutinized by \citet{Tolstoy2009}.
\citet{Mcconnachie2012} has painstakingly assembled a homogenized
database of the properties of all known dwarfs within 3
Mpc. \citet{Walker2012} has written down a scrupulous account of the
current evidence for the presence of Dark Matter in dwarfs, while
\citet{kravtsov2010} has reviewed the progress in reconciling the
mismatch in the appearance of the real dwarf satellites and the toy
ones built into simulated Dark Matter sub-structure. The quest for the
least luminous galaxies, so-called {\it ultra-faint dwarfs}, has been
documented by \citet{Willman2010}. To complement these, a review of
the structure, chemistry and dynamics of the Galactic stellar halo can
be found in \citet{Helmi2008}.  Finally, \citet{Ivezic2012} explain
exactly how the large surveys like the SDSS have revolutionized the
way research into the Galactic stellar populations is conducted.

This review will attempt to avoid boring the reader with re-stating
the facts already discussed thoroughly in the works above. Instead,
its purpose is to report on the most recent progress in the area of
the Galactic Archaeology and to list some of the burning questions
that are destined to be answered with the upcoming sky surveys.

\section{Little galaxies. Big questions.}

The main premise of the current galaxy evolution theory, which itself
exists within the broader theory of the Universal structure formation
a.k.a. $\Lambda$CDM, postulates that all galaxies are born, live and
die inside dark matter (DM) halos. $\Lambda$CDM uses Cold Dark Matter
to provide nucleation sites for the subsequent budding of galaxies of
all sizes. Small lumps in the primordial DM gravy are the most
numerous and develop the quickest, therefore the first baryonic
systems to appear are the dwarf galaxies. What happens later is the
competition between the expansion of the Universe and the
gravitational pull of the emerging DM halos. In this game, the
majority of the dwarfs eventually lose: they are dragged into deeper
potential wells, where they get undone and their matter, dark and
otherwise, is subsumed to form bigger galaxies.

A small fraction of the aboriginal dwarf satellite population that
survives the tidal disruption during the accretion should, in
principle, be detectable in and around the Galaxy today. $\Lambda$CDM
is a young theory and it is perfectly reasonable that its fundamental
properties are still being actively questioned. One key prediction of
the theory is the existence of the large number of small DM halos
(called {\it sub-halos} to indicate the hierarchical nature of all
structures in this paradigm) in a galaxy like our own. The theory then
postulates that dwarf satellites are the sub-halos that have accreted
enough gas and have held on to it long enough to cool it down and form
stars. Those dwarfs that have survived until $z=0$ must have either
stayed well away from the strong tides of the central Milky Way or are
just arriving to their final destination \citep[see
  e.g.][]{Bullock2005}.

\subsection{Galactic dwarf census}

Galactic Archaeology provides the observational evidence of the
accretion onto the Milky Way and the statistics of the survived and
the destroyed dwarfs. This data can be interpreted within the current
hierarchical structure formation model provided the processes to do
with star formation and interaction between the baryons and the DM are
well understood. For example, \citet{Klypin1999} demonstrate how
merely the count of the Galactic dwarf galaxies can be used to
challenge the very foundations of the accepted galaxy formation
paradigm. As they show, it is possible that to reconcile the
measurement and the prediction of the satellite mass function, a
cut-off in the power spectrum of the primordial matter fluctuations on
small scales is required. Alternatively, they point out, it is
perfectly feasible that the suppression of star formation in low-mass
systems is stronger than expected, thus leaving the vast majority of
sub-halos dark forever.

The number of the low-mass galaxies around the Milky Way has more than
doubled in the last ten years solely due to the supply of high quality
all-sky data from the SDSS. Yet, any attempt to corroborate the theory
of dwarf formation based on these recent discoveries, would be
hopeless without first quantifying how much the rapidly growing
satellite sample is influenced by the selection effects.  All Galactic
dwarfs (excluding the Sagittarius and the Sextans dSphs) discovered
before the first SDSS data releases and now conventionally known as
{\it Classical} were found by simply eye-balling the photographic
plates. The SDSS has changed the game completely: today the
science-ready catalogs containing hundreds of millions of stars and
galaxies can be trawled through quickly, making the search efficient
and quantifiable. Using the early released data from only two SDSS
runs, \citet{Willman2002} estimates the sensitivity of the full survey
dataset to resolved Galactic companions and predicts that the
satellite Luminosity Function can be straightforwardly constructed
facilitating the first unbiased comparison between the theory and
observations. Accordingly, \citet{Koposov2008} present the
completeness calculation for the SDSS DR5 based on the fully-automated
satellite detection algorithm and a large suite of realistic mock
observations of dwarf galaxies of various sizes and luminosities. The
results presented by \citet{Koposov2008} show clearly that there are
large swathes of the parameter space where the Milky Way satellites
can be detected with nearly $100\%$ efficiency.  Outside these
regions, the detectability drops sharply to 0. The transition boundary
is controlled by the surface brightness limit of the SDSS $\mu_{\rm
  lim}$, the luminosity of the satellite and its
distance. Alternatively, for all objects with $\mu_{\rm lim} < 30$ mag
arcsec$^{-2}$, the SDSS completeness can be expressed in terms of the
volume within which it discovered {\it all} satellites of a particular
luminosity.

\begin{figure}
\centering
\includegraphics[width=0.95\linewidth]{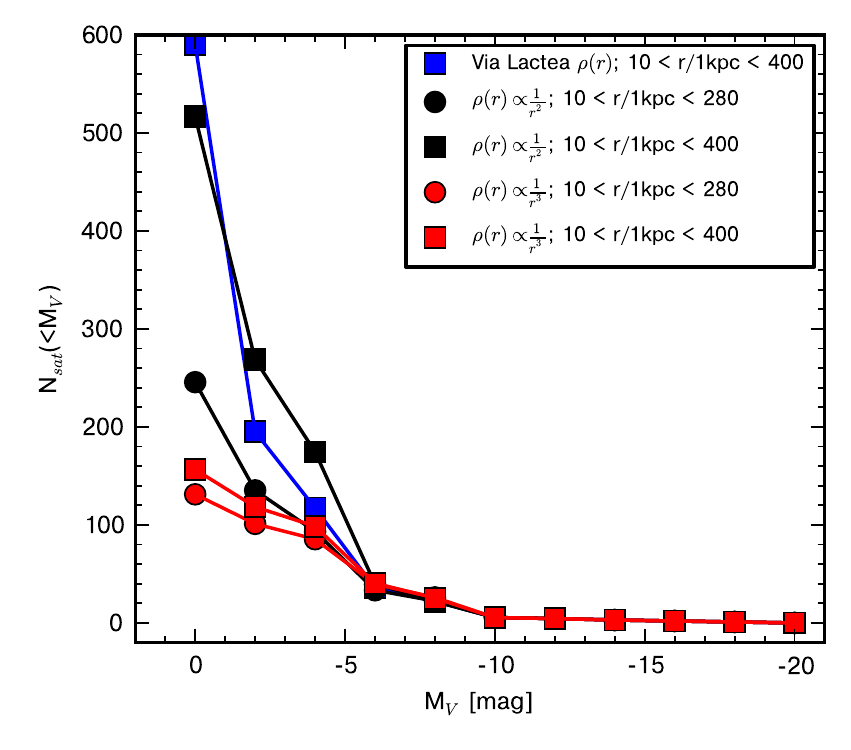}
\caption{Dependence of the Galactic dwarf luminosity function on the
  assumed radial density profile of the satellites. This shows the
  cumulative luminosity functions (the total number of objects as a
  function of their absolute magnitude) for the Milky Way satellites
  in the spherical volume with radius $R=280$ kpc (filled circles),
  which is similar to the volume used in \citet{Koposov2008}; and with
  $R=400$ kpc (squares), similar to the maximal distance used in
  \citet{Tollerud2008}. Red curves give the LFs obtained with the
  satellite radial profile proportional to $r^{-3}$, while the LFs
  given in black assume that the number of satellites decays as
  $r^{-2}$. Blue curve is the LF based on the radial distribution of
  the Via Lactea sub-halos. All density laws are truncated at $R=10$
  kpc, the satellite detection efficiency applied is from
  \citet{Koposov2008}. Only $\sim 100$ dwarfs are predicted to inhabit
  the Galaxy if their radial number density follows the $r^{-3}$ law
  (red curves), i.e. a NFW distribution with small scale radius, see
  also \citet{Koposov2008}. Substantially more ultra-faint objects are
  anticipated if the inverse square law is adopted, particularly if
  the dwarf population extends to distances as large as $R=400$ kpc
  (black squares). In Via Lactea simulation, in the inner 60-80 kpc,
  the number density of sub-halos is even flatter as shown in the
  Supplementary Figure 1 of \citet{Diemand2008}. Accordingly, most of
  the faintest satellites are supposed to be undetected by the SDSS,
  as reflected in the quick rise of the LF (in blue) at $M_V \sim -3$,
  see also \citet{Tollerud2008}. Figure courtesy of Sergey Koposov,
  IoA.}
\label{fig:lf}
\end{figure}

According to \citet{Koposov2008}, the sample of dwarf satellites with
luminosities brighter than $M_{V}\sim -5$ is essentially complete out
to the Galactic virial radius $r_{\rm vir}=280$ kpc within the SDSS
DR5 field of view. However, the accessible volume plummets fast with
decreasing dwarf luminosity, which means that for satellites as faint
as Segue 1 with $M_V\sim-3$, only few percent of the virial volume
have been probed. Importantly, using thus-calculated fraction of the
total Galactic volume sampled, it is now possible to predict the
complete number of the satellites if their distribution with radius is
assumed. Naturally, the flatter the radial distribution of the
satellites the bigger is faction of objects remaining to be
detected. \citet{Koposov2008} surmise that if the number of dwarfs
decays in a NFW-like fashion ($\propto r^{-3}$ at large distances)
then the Luminosity Function (LF) of the Milky Way satellites goes as
$\sim 10^{0.1(M_V+5)}$. According to this LF, the objects with
luminosities between $M_V=-2$ and $M_V=-5$ contribute just under a
half of the total $\sim 100$ dwarfs. Some 4 times more satellites is
predicted by \citet{Tollerud2008} who take advantage of the
completeness calculation published by \citet{Koposov2008}, but choose
to adopt the radial distribution of sub-halos from the Via Lactea
N-body simulation \citep{Diemand2008}. In the inner 60-80 kpc, the
distribution of sub-halos in Via Lactea drops very slowly with radius.
In fact, for several tens of kpc it is almost flat according to the
Supplementary Figure 1 of \citet{Diemand2008}. The effect of this
large ``core'' in the satellite number density profile on the dwarf LF
is particularly striking for the faintest of the satellites, those
detectable by the SDSS only out to 30-40 kpc. Out of the fiducial
$\sim$ 400 satellites predicted by \citet{Tollerud2008}, almost $90
\%$ are those with $-2 < M_V < -5$. The drastic dependence of the
Galactic dwarf LF on the assumed radial distribution of satellites is
illustrated in Figure~\ref{fig:lf}. From these first attempts to gauge
the size of the Galactic satellite body, the important role played by
the faintest of the dwarfs emerges.

\subsection{Unexplored variety of hierarchical galaxy formation}

In the decade following the publication of the two whistle-blowing
papers by \citet{Klypin1999} and \citet{Moore1999}, the {\it sub-halo
  abundance matching} (SHAM) technique that relies on assigning higher
stellar luminosities to the DM sub-halos with higher maximal masses,
has been optimized near to perfection. The success of the abundance
matching has created a brief period of the comforting lull with the
gaping void between dwarf galaxies and sub-halos (known as the {\it
  missing satellites problem}) seemingly breached
\citep[e.g.][]{Koposov2009,Maccio2010}. However, pitted against the
existent Galactic satellites, these mock dwarfs do not stand a close
scrutiny: the density profiles in the most massive systems are not
corroborated by the available kinematics \citep[see
  e.g.][]{Boylan-Kolchin2012}. It turns out, the overall
star-formation efficiency is now such a strong function of the
sub-halo mass that the dwarfs with modest stellar masses and velocity
dispersions are forced to inhabit disproportionally massive and dense
halos. What is missing from this picture drawn with the help of SHAM?
It appears quite a few of the vital ingredients might be lacking.

To begin with, the influence of the rather significant baryonic disk
in the Galaxy has been surprisingly overlooked in many of the SHAM
studies. Yet, as demonstrated in a recent succession of papers
\citep[e.g.][]{Taylor2001,Read2006,Penarrubia2010,D'onghia2010}, disks
aide the tidal destruction of satellites thus seriously depleting the
number of the dwarf survivors. One crucial detail is noted by
\citet{Penarrubia2010}: dwarf galaxies with cored density profiles are
less likely to survive the devastating action of the
disk. $\Lambda$CDM on its own does not permit cored dark matter
profiles, but \citet{Read2006} and \citet{Pontzen2012} show that the
gas flow due to the persistent supernova feedback can pull dark matter
along to the outskirts of dwarf galaxies. Accordingly, a plausible
setup in which the satellite survival rate at $z=0$ is regulated by
the interplay between the strong stellar feedback evacuating the
centers of sub-halos and the enhanced tidal disruption due to the disk
is described in the work by \citet{Brooks2013}. Here, by means of
applying a simple correction to the central masses of semi-analytical
dwarfs, as originally proposed by \citet{Zolotov2012}, many of the
massive satellite galaxies are wrecked, and the tension between the
data and the theory seems to be alleviated once again. Note, however,
that the total amounts of supernova energy required to cause
appreciable damage to the DM central density cusps have been deemed
excessive by many authors \citep[e.g.][]{penarrubia2012,gk2013}.

As the physics of star formation is just starting to be explored,
there does not exist a single hydro-dynamical simulation of the Milky
Way run at the resolution appropriate to resolve the gas infall and
cooling at all epochs from high redshifts to the present day. Instead,
on a star by star basis, the processes that play the most important
role (like cooling and feedback) are gleaned, synopsized and
subsequently incorporated into the simulations as {\it sub-grid}
recipes to be followed together with the laws of the Newtonian gravity
(and sometimes hydrodynamics). In this approach, it is believed that,
the evolution of the dark matter density on the relevant scales has
been fully captured with the latest pure N-body simulations. However,
as the DM particles are followed from the distant past to the current
day, the actual sequence of accretion events the Milky Way prototype
goes through varies considerably from host to host. This has profound
implications on the final shape of the DM sub-halo mass function (MF),
and ultimately on the properties of the Galaxy's dwarf satellite
population.

\subsubsection{Host mass and concentration}

When faced with the myriad of DM halos to furnish a Milky Way together
with its satellites, the conventional choice is to select the host by
matching its virial mass to that of the Galaxy. Note that according to
Figure 8 in \citet{Springel2008}, in the 6 host halos with different
virial masses, the sub-halo number counts in the bins of mass scaled
to the the mass of the host lie exactly on top of each
other. Accordingly, it is established that the relative MF scales as
$(M_{sub}/M_{50})^{-1.9}$ and the absolute sub-halo abundance
normalization includes another factor of $M_{50}$ for the host
mass. Thus, for the hosts whose masses are different by $100\%$ the
total sub-halo counts would also disagree by a factor of 2. Alas, the
mass of the real Milky Way is not known with the accuracy as high as
$100\%$. Even though, depending on the method used, the formal
uncertainties can be as low as $30\%$, there are serious disagreements
between the measurements making the systematic error much higher. For
example, according to \citet{Watkins2010}, the plausible range for the
Milky Way mass within 300 kpc is approximately from $1 \times 10^{12}
M_{\odot}$ to $3 \times 10^{12} M_{\odot}$, which would imply a factor
of $\sim$3 difference in the total number of sub-halos.

The present uncertainty in the measurements of the Milky Way's
concentration is $\sim 100\%$, which is perhaps even more appalling
since the allowed range of concentrations is much smaller. The
concentration $c=r_{vir}/r_s$ of a DM halo describes how dense its
inner parts (within the scale radius $r_s$) are compared to the halo
overall (out to the virial radius $r_{vir})$. For the halos of similar
virial mass, their concentrations are ultimately linked to the shape
of the host's mass assembly history, with those peaking at very early
times producing higher concentrations \citep[see
  e.g.][]{Wechsler2002}. For the population of dwarf satellites at
redshift zero, having a host halo with a high concentration is a
double whammy. The first implication is obvious: if the accretion
history settled into the quiescent phase at high redshifts, at later
times, fewer dwarfs will be accreted. Second, having a dense pile up
of dark matter (and baryons) means more efficient tidal disruption and
the lower satellite survival rate. According to the N-body
simulations, for the halos with Milky Way-like masses, the
concentration is predicted to be of order of $c=12 \pm 3$ \citep[see
  e.g.][]{Boylan-Kolchin2010}.

There are four techniques available today to measure the Galactic
mass, each with its own assumptions and inherent limitations. The
first three rely on the kinematics of a sample of the gravitational
potential tracers, and therefore can only be applied straightforwardly
within few tenths of the Galactic virial radius. The first method uses
stars or gas to determine the run of the Galactic rotation velocity
$v_{\rm rot}=\sqrt{r\frac{{\rm d}\Phi}{{\rm d}r}}$ (where $\Phi$ is
the underlying potential) with Galacto-centric radius $r$. Naturally,
the rotation curve can only be sampled within $r < 20$ kpc, as there
is no indication that the disk continues much beyond that. Having the
full 6D information is rare, the latest such attempt presents the
measurements of trigonometric parallaxes for a dozen or so masers,
from which the circular rotation speed at the Sun's location is
deduced \citep{Reid2009}.  Alternatively, \citet{Bovy2012} shows that
the circular velocity can be inferred by marginalizing over poorly
constrained distances and unknown proper motions for a large
($>3,000$) set of (mostly) disk giant stars with accurate
line-of-sight velocities.

An equally rewarding, but perhaps yet a more challenging approach is
to gauge the Galactic escape speed $v_{\rm esc}=\sqrt{2|\Phi|}$ by
analyzing the tail of the stellar velocity distribution. The results
are sensitive to the quality of the distance and the proper motion
data, in particular, imperfect proper motion measurements are so
detrimental that, normally, they are avoided altogether. Instead, a
velocity distribution function is chosen, whose exact shape is
controlled by a small number of parameters that get simultaneously
constrained in the process of the likelihood maximization. For
example, using a relatively small sample (16) of high-velocity stars
provided by the earlier releases of the RAVE survey, \citet{Smith2007}
measure the local escape speed. Conveniently, given the Galactic
escape speed and assuming the contributions of the bulge and the disk
to the total potential, the mass and the concentration of the Milky
Way's halo also can be extracted. The analysis by \citet{Smith2007}
seems to prefer the Galaxy with the mass as low as $0.9 \times 10^{12}
M_{\odot}$ and the concentration as high as 24. While the
applicability of both the circular speed and the escape speed
techniques is restricted to the inner Galaxy, the latter has the
advantage of probing the Galactic mass out of the disk plane.

Most of the Milky Way's mass lies beyond the extent of the disk, hence
at large Galacto-centric distances, a different approach is required.
Given enough mass tracers (stars or satellites) in a wide range of
locations throughout the Galaxy, the total mass profile can be
obtained by means of Jeans modelling of the tracer kinematics
\citep[see e.g.][]{Battaglia2005}. The terms that enter the spherical
Jeans equation are: the tracer density, the tracer velocity dispersion
and the tracer velocity anisotropy. At large distances, only one of
these might be available, namely the line-of-sight velocity
dispersion. Making the Jeans analysis of the far reaches of the
Galactic halo possible clearly falls within the realm of Galactic
Archaeology which can both deliver the most distant tracers as well as
constrain the overall tracer density distribution.  The stumbling
block, however, is the scarcity of tracers with the tangential
components of the velocity measured. As a consequence, the anistropy
is normally treated as a nuisance parameter since the most datasets
available lack in accuracy and breadth to constrain it. Even with the
arrival of Gaia, the situation will only improve for the nearby
objects, leaving the distant ones wanting in more precise proper
motions. While assigning anistropy to a tracer population is a
solution far from ideal, presently, it is the Jeans modelling together
with its variants that provides the most stringent constraints on the
total mass of the Milky Way \citep[e.g.][]{Xue2008}.

Finally, a new, conceptually different method to probe the matter
distribution in the Galaxy is now coming of age. Compared to the three
approaches discussed above, it does not rely at all on the
instantaneous kinematic properties of large samples of tracers, and
thus, for example, needs no assumption of their velocity
anisotropy. Stellar {\it tidal streams} are shown to align closely
with the obit of their disrupting (or disrupted) progenitor and
therefore give an almost direct way of measuring the underlying
potential. Recently, the power of the method has been demonstrated
beautifully by \citet{Koposov2010} who, using the 6D data of the GD-1
stream, measured the Galactic rotation curve locally. This type of
analysis can, in principle, be extended to distances beyond the
predicted Galaxy's scale radius $r_s$. The prime source of degeneracy
in recovering the Galactic potential using tidal tails, is the length
of the stream available. However, to date, for several distant streams
there exists sufficient data covering tens \citep[Orphan Stream with
  the maximal distance of $\sim 50$ kpc][]{Belokurov2007a,
  Newberg2010} or even hundreds of degrees \citep[Sagittarius Stream
  with the maximal distance of $\sim 100$ kpc, e.g.][]{ Majewski2004,
  Newberg2003,Belokurov2006b, Yanny2009, Belokurov2013}. Given the
magnitude limit of the on-going imaging surveys like SDSS or
Pan-STARRS, for stellar streams to be detected so far out in the halo,
the progenitor's luminosity, and therefore mass, ought to be
substantial. This bias implies that the currently known distant
streams can not be appropriately modeled using simple orbit
approximation, the circumstance that now can be mitigated with the
arrival of more sophisticated modeling techniques
\citep[e.g.][]{Eyre2011,Sanders2013}

\subsubsection{Mass assembly history and environment}

The computational expense of running numerical simulations of Galactic
halos at the resolution adequate to capture the properties of the halo
sub-structure is prohibitively high. Hence, the comparison between DM
sub-halos and the observed dwarfs has been based on the analysis of
only 8 N-body simulations: a sample of 6 Aquarius halos
\citep{Springel2008}, complemented by halos of Via Lactea II
\citep{Diemand2008} and GHalo \citep{Stadel2009}. For this reason, the
host-to-host variation of the dark and the luminous sub-structure
remains largely un-studied. As well as improving the resolution and
the speed of the simulations, there is an ongoing effort to quantify
the complex diversity of structures forming within $\Lambda$CDM with a
handful of key parameters, e.g. host halo mass, shape of the accretion
history and significance of the overdensity of the local
volume. These, of course, are inter-related: the mass of the DM halo
hosting a Milky Way galaxy at redshift $z=0$ is the sum total over its
accretion history, which in turn is dictated by the whereabouts of the
halo within the cosmic Large Scale Structure. While the importance of
not knowing such an elementary property of the Galaxy like its mass is
now accepted, the impact of the location of the Milky Way within the
larger cosmic structure and the details of its accretion history are
just beginning to be investigated.

Today, there exist two intriguing constraints on the Milky Way's
accretion history.  First is the observation that the Galactic disk
probably has to survive intact for some 7-10 Gyr \citep[e.g. Figure 18
  of][]{Burnett2011}. This, therefore, potentially excludes any
significant mergers between $z \sim 1$ and now. Second is the new
observational and numerical evidence for the late infall of the
Magellanic Clouds \citep[e.g.][]{Besla2010}. This signifies the end of
the quiescent phase in the Galactic accretion history and can be
exploited to place useful constraints on the mass assembly of the
Galaxy \citep[e.g.][]{Busha2011}. What happened before the quiescent
phase, why did it begin and why did it end?  How common is this
particular shape of the {\it mass assembly history} (MAH) amongst
other disk galaxies of similar mass?  Was the early accretion
dominated by small satellite infall and was it synchronized?  Or
perhaps, was the bulk of the Galactic matter instead acquired in one
or two mergers with massive nearby fragments?  Unfortunately, these
questions remain largely unanswered and therefore, a variety of loose
ends continues to confuse the current picture of the Galaxy formation
and muddle the modelling of the nearby dwarfs. For example, if many
small satellites are accreted early on, enough should survive and be
detectable today. On the contrary, massive mergers usually lead to an
entirely different outcome: in this case, the dynamical friction is
strong enough to slow the dwarf down thus boosting its plunge into the
inner Galaxy where it is quickly disrupted. These two scenarios can be
identical in terms of the epoch of accretion and the total mass
accreted, yet they can produce dramatically different dwarf satellite
populations at $z=0$.

An attempt to quantify the amplitude of the host-to-host scatter in
the properties of artificial Galactic dwarfs using analytic models is
presented in \citet{Purcell2012}. The p\`iece-de-r\'esistance of the≠≠≠≠
method is the Monte-Carlo sampling of an arbitrary large number of
different accretion histories \citep[as described
  in][]{Zentner2005b}. Using this technique, it can be demonstrated
that the scatter in the possible MAHs is naturally large enough for
the Milky Way-like halo to host a satellite population consistent with
the observed one in 10\%-20\% of cases. These results, within the
limitations of the method, shed light onto the statistical
significance of the ``too-big-to-fail'' problem
\citep{Boylan-Kolchin2012}: there does not have to be a serious excess
of massive invisible sub-halos in the Galaxy. Interestingly, together
with the recently invoked lower Galaxy mass
\citep[e.g.][]{Vera-Ciro2013} and the strong stellar feedback
\citep[e.g.][]{Brooks2013}, this is now the third solution for the
potential problem identified by \citet{Boylan-Kolchin2012}. It would
seem that if all three methods are as efficient as described, there
could be very few satellites left around the Galaxy! It is, therefore,
the most urgent task for the Galactic Archaeology to provide new
observational constraints of the Milky Way's accretion history through
studies of the spatial and the chemo-dynamical distributions of the
ancient stellar halo populations.

The Milky Way is not a solitary field spiral: together with its
neighbor of approximately the same mass, Andromeda and its satellites,
it makes up the small slightly over-dense region of the Universe known
as the Local Group of galaxies. The so-called {\it assembly bias}
stipulates an excess of probability of finding a massive satellite
sub-halo around hosts situated in higher density regions as compared
to those in under-dense environments
\citep[e.g.][]{Wechsler2006}. Possibly, this effect could go some way
to explaining the presence around the Milky Way satellites as massive
as the Magellanic Clouds. According to \citet{Busha2011b}, while for
the field halo of Milky Way-like mass, the probability to host LMC/SMC
pair is of order of $5\%$-11$\%$, having another host halo of similar
mass in the vicinity boosts it up to 25$\%$. This is good news, but
are these sub-halos on their first (or perhaps second) passage around
the simulated Galaxies as the Milky Way observations seem to indicate?
A unique investigation is described in \citet{Forero-Romero2011} who
use a suite of so-called constrained simulations of the Local Group
(CLUES, see http://www.clues-project.org/) in which the broad-brush
features of the Milky Way-Andromeda pair are reproduced, to study the
assembly history of either host halo. They find that i) both galaxies
had their last significant accretion event some 10-12 Gyr ago, and
that ii) this particular common accretion history is quite rare (from
1$\%$ to 3$\%$) amongst the pairs of host halos in Bolshoi
simulation. This conclusion appears to be in contradiction with the
studies in which the Clouds are just being accreted.

\subsection{Tidal origin of the local dwarf galaxies}

It is inspiriting that there exists at least one alternative, and,
importantly, testable scenario of the formation of dwarf satellites in
and around the Milky Way. \cite{Lynden-Bell1976} first pointed out the
proximity of the several of the Galactic dwarfs to the LMC's orbital
plane as defined by the gaseous stream leading the Cloud. The
hypothesis then put forward is of a Greater Magellanic Galaxy that had
been torn apart as it interacted with the Milky Way, giving birth to
the Large and Small Clouds, as well as to a litter of smaller
dwarfs. A quarter of a century later, with the measurement of the
space velocities of the satellites in hand, the surprising
juxtaposition of the orbital planes of the LMC, SMC, UMi and Dra is
confirmed \citep[e.g.][]{Palma2002}. This motivates \citet{Kroupa2005}
to claim that the observed distribution of the Galactic satellites is
too anisotropic to fit seamlessly within the CDM paradigm. In the
authors' opinion, such alignment (dubbed later as the ``disk of
satellites'', DoS) is prohibitively rare in computer simulations of
galaxy formation in the Universe full of Dark Matter: the accreted
sub-halos should have had enough time to relax in the Milky Way's
potential, thus erasing any signs of coherence.

It is, however, certainly too naive to believe that in $\Lambda$CDM
Universe, the distribution of dwarf satellites around a Milky Way-like
host is always isotropic. \citet{Zentner2005} show that through the
combined effect of i) filamentary accretion and ii) the alignment of
sub-halo orbits with the major axis of the triaxial host halo, the
probability of choosing the simulated sub-halo populations from an
isotropic distribution is as low as $10^{-4}$. The success of these
simulations in assembling anisotropic satellite distributions is
curious since these particular host galaxies do not posses disks. The
presence of a baryonic disk should help to get rid of the satellites
orbiting near it, thus making the satellite distribution more
anisotropic. \cite{Libeskind2005} use a slightly different numerical
setup to generate their host halos as well as their satellite
galaxies but come to the same conclusion: a good fraction of the
brightest satellites is bound to end up in a plane-like arrangement
having arrived to the host through 1 or 2 primary filaments.

While \cite{Lynden-Bell1976} only briefly mentions a possible scenario
in which the parent galaxy dissolves to leave several smaller
fragments behind to be observed today as dwarf satellites,
\citet{Kroupa2005} go further to suggest the exact mechanism
responsible for their production. They speculate that the creation and
the subsequent compression of the gaseous tidal tails is followed by
tail fragmentation and active star-formation. It is claimed that the
stellar systems born in this violent process, also known as {\it tidal
  dwarf galaxies} can survive long enough. If they do, their
anisotropic distribution on the sky is merely the consequence of the
proximity of their birthplaces in the tidal tail that is now
vanished. This dSph formation mechanism advocated by
\citet{Kroupa2005} harks back to their earlier dynamical work
\citep{Kroupa1997}, in which a quasi-stable solution for a dSph-like
DM-free stellar system is discovered. With the help of a suite of
simple N-body simulations, it is argued that a tidal dwarf galaxy in
the last throws of disruption can posses apparent surface brightness
and velocity dispersion not unlike those observed in dSphs around the
Milky Way. As \citet{Kroupa1997} argues such high velocity dispersions
would lead to over-estimated masses and therefore to highly inflated
mass-to-light ratios, while the actual $M/L$ remains quite
low. \citet{Metz2007} re-run the experiment and show that their
simulated tidal dwarf remnants and the Galactic dwarfs can look alike,
especially within the region of the structural parameter space
occupied by the ultra-faint satellites. Even though the fact of the
existence of such out-of-equilibrium satellite configurations in
numerical simulations is established, as of today, no evidence has
been found that they can persevere for longer than a 1-2 Gyrs
\citep[see e.g.][]{Casas2012}.

As the census of the sub-structure in the halos of the Milky Way and
the Andromeda galaxies is being filled in fast, the growing sample of
satellites and streams allows for more rigorous tests of possible
anistropies in their spatial and kinematic distributions. For example,
\citet{Pawlowski2012} extend the study of the Galactic ``disk of
satellites'' to include the known stellar and gaseous streams. Their
argument in support of the previously found DoS orientation is that 7
out the 14 streams they analyse align well with the disk. With this
observation in hand, they claim that it is not merely the ``disk of
satellites'' that surrounds the Milky Way, but rather a ``vast polar
structure'' (VPOS) appears to dominate the Galactic sub-system
distribution at all radii. Once again the conclusion is reached that
the presence of such structures is in contradiction with the $\Lambda
CDM$ theory. Before the probability of encountering this so-called
VPOS is worked out for the current galaxy formation paradigm, it is
worth noting that while the number of the streams contributing to it
seems large (a half of the total considered), their combined mass is
minuscule. Therefore, these (in particular stellar) streams contribute
close to nothing to the significance of the supposed anisotropy in the
Galactic halo.

Curiously, in the case of the M31, \citet{Ibata2013} exhibit plausible
evidence for the planar alignment of nearly half of the dwarf
satellites. Moreover, these appear to be co-rotating around Andromeda
in a semblance of a disk, which contains the line connecting the host
galaxy and the Milky Way. This discovery is responsible for another
attempt to debunk $\Lambda CDM$ this time by \citet{Hammer2013} who
develop their earlier idea of a major merger at the M31 location
\citep[see e.g.][]{Hammer2007} and suggest that most of the dwarf
galaxies, including the Magellanic Clouds have formed as a result of
this upheaval. 

Overall, it seems that the hypothesis in which dwarf satellites are
born in major merger events can give a convincing account of the
observed distribution of satellites on the sky. However, currently the
theory does not stack up against the entirety of the observational
evidence, both locally (e.g. the extended star-formation histories and
the extremely old stellar populations of the Milky Way dwarfs) as well
as outside the Galaxy (e.g. low major merger rates for L$_*$ hosts).

\section{Archaeologist's toolbox}

\begin{figure}
\centering
\includegraphics[width=0.93\linewidth]{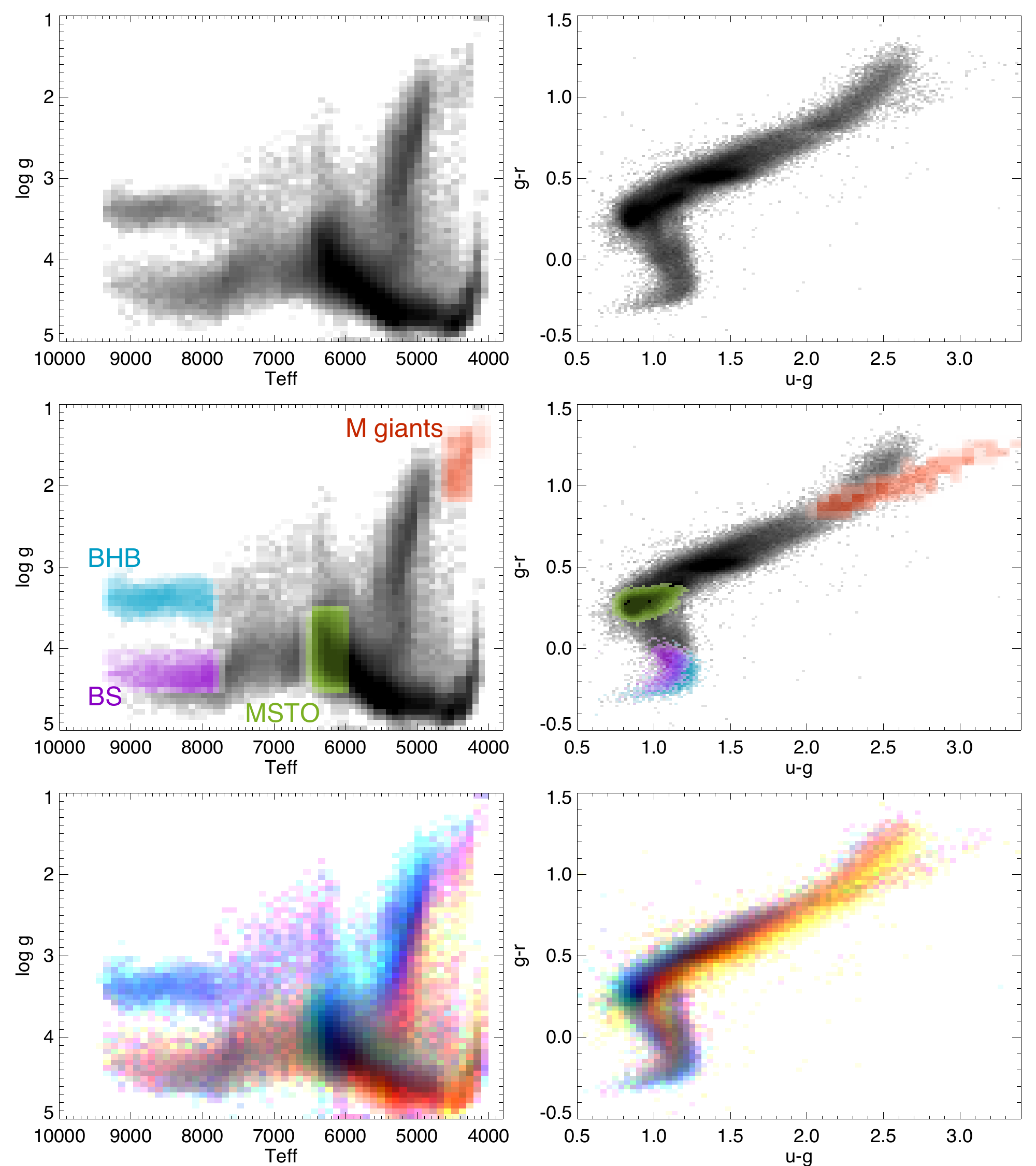}
\caption{Stellar tracer selection in the SDSS database. {\it Left:}
  Density of stars in the plane of surface gravity $\log g$ and
  effective temperature $T_{\rm eff}$ for $\sim 180,000$ DR8 spectra
  with $15 < g < 17.5$. {\it Right:} Stars with spectroscopy from the
  left column are plotted on the plane of $u-g$ and $g-r$ color. {\bf
    Top:} overview of the sample, darker shades of grey indicate
  higher density. {\bf Middle:} Selecting the tracers. BHB (blue),
  Blue Straggler (violet), MSTO (green) and M-giant (red) stars are
  chosen in the left column based on their temperature and surface
  gravity. Density of selected stars is then over-plotted in $u-g$,
  $g-r$ space using the same color scheme. {\bf Bottom:} Metallicity
  distribution in the sample. This shows false RGB images (left and
  right) constructed with 3 grey-scale density distributions of stars
  picked based on their $[Fe/H]$. Red component is for metal-rich stars
  with $-0.75 <[Fe/H]< 0$, green (intermediate) $-1.5 <[Fe/H] <
  -0.75$ and blue (metal-poor) $-3 <[Fe/H] <
  -1.5.$} \label{fig:tracers}
\end{figure}

Low-mass stars (around $\sim 1 M_{\odot}$) shine for billions of
years, and therefore keep the record of historical events in the Milky
Way. To be able to read into the Galactic diary, collections of stars
with comparable chemistry, age or, at least, similar luminosity class
must be identified. The distributions of such {\it stellar tracers} in
two (positions on the sky), three (place on the sky and along the line
of sight), four (location in space and in radial velocity) or even
seven (configuration space and velocity space coordinates together
with chemistry) dimensions are then measured to benchmark, with some
help from Galactic Dynamics, the theories of structure formation.

The Galaxy endlessly churns the pieces of smaller satellites it
acquires, continuously smoothing the spatial densities of the debris.
The rate at which the Galactic blender operates decreases from the
centre outwards. Far out in the halo, where the orbital periods reach
giga-years, unbound stellar sub-structures can maintain superficial
spatial coherence for eons. However, closer to the Solar radius, extra
(dynamical or chemical) information is required to filter out
particular debris from the smooth mess. Therefore, the interplay
between the number of useful stellar tracers, the information content
per star, and the overall volume probed is what determines the
relevance of a Galactic halo survey.

In the not-so-distant future, with the data from the Gaia astrometric
space mission and a host of planned large-area spectroscopic surveys,
it should be possible to paint the unambiguous picture of the events
that took place in the Galaxy between redshift $z=20$ and redshift
$z=0$. At the moment, we will have to make do with what we have
got. The observational advances in Galactic Archaeology made in the
last few years happened thanks to a handful of wide area imaging
surveys, namely 2MASS and SDSS, and massive spectroscopic efforts such
as Segue and RAVE.

Of the several sky surveys of past decade, the SDSS appears to have
been operating in a sweet spot: it turns out a 54 second exposure is
long enough to reach Main Sequence stars at distances of several tens
of kpc from the Sun, and thus yield an unprecedented 100 million
object database; yet short enough to see plenty of the sky in limited
amount of time. The now classic $ugriz$ filter set encodes the stellar
spectral energy distribution (SED) into a compact form, but preserves
enough frequency diversity to study in detail a variety of stellar
populations. This section therefore mostly concentrates on the
observed properties of the Galactic stellar halo as seen by the SDSS
(and its extensions) outside the Solar radius.

\subsection{Stellar tracers of the Galactic halo in the SDSS}

\begin{figure}
\centering
\includegraphics[width=0.99\linewidth]{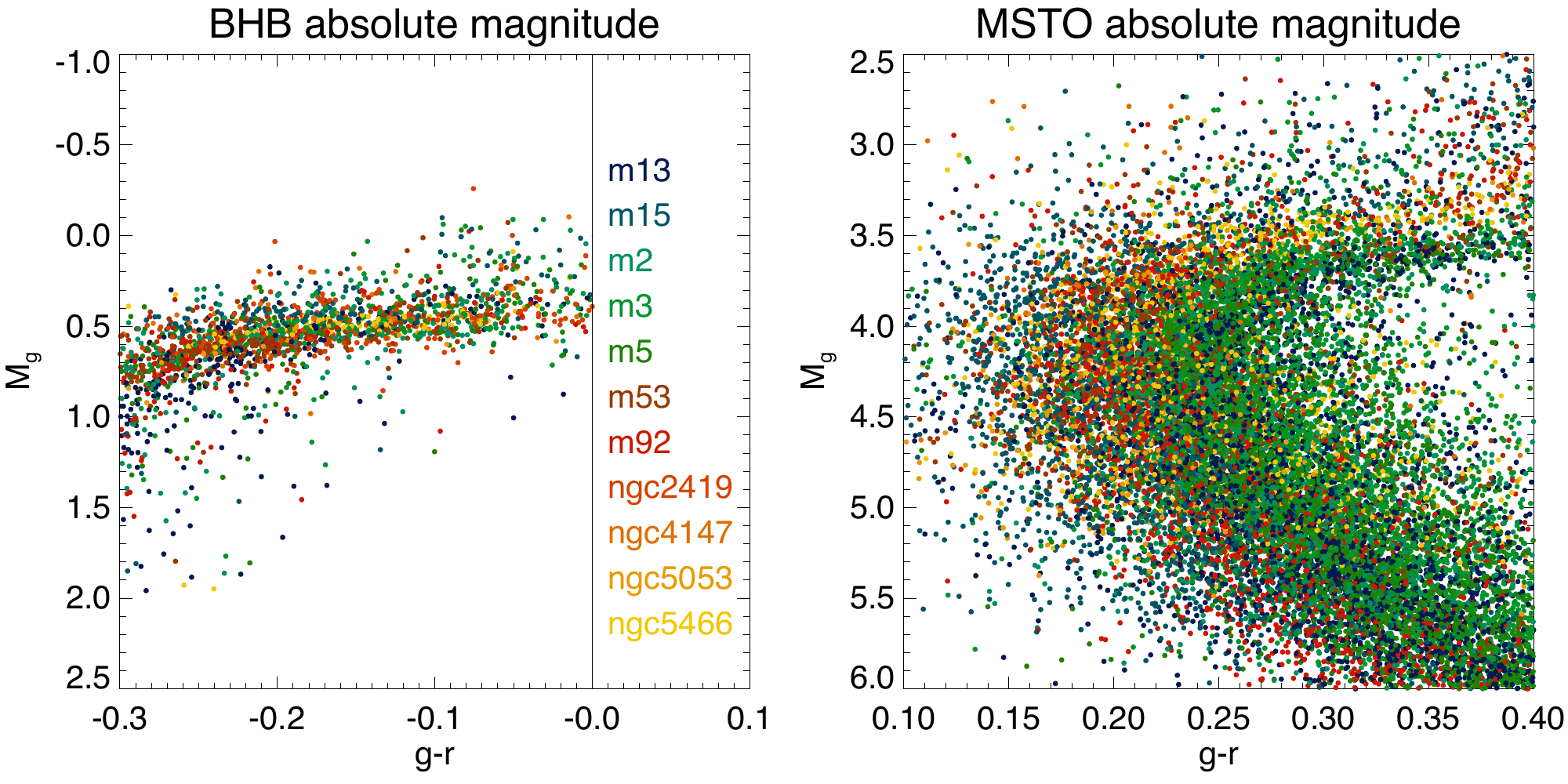}
\caption{Absolute magnitude of stellar tracers. {\it Left:} Blue
  Horizontal Branch star candidates in 11 Galactic star clusters. Each
  dot represents one BHB, stars from different clusters are marked
  with different color. Cluster name and the color convention are
  shown in the inset. Once a model for the slight variation of the
  luminosity with color has been applied, the absolute magnitude of a
  BHB star can be estimated with accuracy $\lesssim 0.1$ mag. {\it
    Right} Stars with $g-r < 0.4$ in 11 Galactic star clusters. Apart
  from the variation by $\pm 0.5$ mag around the mean magnitude of the
  turn-off $M_g\sim 4$ due to age and metallicity differences between
  clusters, stars on the MS with lower luminosity as well as Sub-giant
  stars bright with higher luminosity are picked up by this $g-r$
  cut. This results in the overall asymmetric spread of $\sim 3$ mag
  in $M_g$.}
\label{fig:bhb_msto}
\end{figure}

There are at least three species of stellar tracers available in the
SDSS photometric data that a Galactic archaeologist can put to
work. In order of decreasing population size, increasing luminosity
and decreasing contamination, these are: Main Sequence Turn Off (MSTO)
stars, Blue Horizontal Branch (BHB) stars and M giant
stars. Figure~\ref{fig:tracers} gives the whereabouts of each of these
three in the space of stellar atmosphere parameters and the space of
broad-band colors.

The left column of the Figure shows the logarithm of density of a
sample of bright ($15<g<17.5$) stars in the spectroscopic Data Release
8 of the SDSS \citep{Aihara2011} on the plane of surface gravity $\log
g$ and effective temperature $T_{\rm eff}$.  The density distribution
of stars in this analog of the familiar Hertzsprung-Russell diagram is
dominated by the pitch-black ribbon of the Main Sequence (MS),
covering the range of $6500 {\rm K} < T_{\rm eff} < 4500 {\rm K}$. The
sharp edge to the MS feature at high temperatures corresponds to the
MS turn-off - these are the brightest of the MS stars and so, ideal
for the halo exploration. The two faint and fuzzy clouds at
temperatures above 7000 K are the helium burning Blue Horizontal
Branch stars, with $3.7 < \log g < 3$, and the ``reinvigorated''
hydrogen burning pseudo-MS stars also known as Blue Stragglers, with
$4 < \log g < 5$. Finally, the cool and inflated stars populate the
Red Giant Branch, attached to the MS at around $T_{\rm eff} \sim 6,500
{\rm K}$ and reaching as high up as $\log g \sim 1.5$. There, right at
the tip sits the small group of M giants, with $T_{\rm eff} < 5000$
K. To guide the eye, the stellar populations mentioned above are
marked in color in the middle panel of the Figure.

As of DR8, only $\sim$0.2\% of all detected stars have been targeted
with SDSS spectroscopy. Therefore, to make surveying the Galaxy's halo
practical, stellar tracers need to be identified by means of
broad-band photometry only. To illustrate the photometric selection,
the right column of Figure~\ref{fig:tracers} gives the logarithm of
the stellar density in the color-color space of $u-g$, $g-r$. Exactly
the same bright stars, those with SDSS spectra, as used for the
creation of the left column are plotted here. As expected, the
behavior of the broad-band color distribution is to do with the
locally measured slope of the SED, and hence is driven by the stars'
temperature. Dwarf and giant stars are not easily separable anymore as
they collapse to form one stellar locus, running from $g-r \sim 0.3$
to $g-r \sim 1.3$. However, it transpires that around the corners of
the locus, the familiar populations can be picked up with ease.

The MSTO stars, being the hottest denizens of the MS, are clumped
right at the blue edge of the $g-r$ distribution as evidenced by the
tight green cluster in the right middle panel of the Figure. It is
obvious that a simple $g-r< 0.4$ cut would produce a relatively clean
sample of the MSTO tracers.  Still bluer in $g-r$, deviating downwards
from the MSTO corner of the stellar locus, lies a ``claw'' of
A-colored but old stars, BHBs and BSs. On further look, following
their loci (marked in blue and violet) as they curve in $u-g$, $g-r$
space, some overlap between the two populations is visible, but more
importantly, in $u-g$ color primarily, the BHB and the BS ridge-lines
stand separated by some 0.15 mag. \citet{yanny2000} provide an
efficient $u-g$,$g-r$ cut which yields a sample of BHB tracers with
minimal contamination from BS or MSTO stars. Unlike the ubiquitous
MSTO stars, the BHBs are manifestations of an old and metal-poor
stellar population, as represented by their abundance in the Galactic
globular clusters.

Even though telling a dwarf star from a giant star photometrically is
pretty much impossible across a wide range of SDSS color, luminous and
metal rich M giants stars peel away and redward from the stellar locus
at around $u-g \sim 2.5$ as the red streak in the right middle panel
of Figure~\ref{fig:tracers} indicates. Equation 1 in \citet{Yanny2009}
stipulates the M giant selection boundaries in the SDSS
colors. Interestingly, age-wise M giants provide a probe of the halo
complementary to that offered by old MSTO and BHB stars. As shown by
\citet{Bellazzini2006} the stars ages range between 5 and 9.5 Gyr,
with an average age of 8 Gyr.

\subsection{Chemical abundances of the SDSS stellar halo tracers}

What are the metallicity biases induced by the particular choice of
the stellar tracers described above? The bottom row of the
Figure~\ref{fig:tracers} shows the metallicity $[Fe/H]$ distribution
of the bright SDSS DR8 stars with spectra. Stars are split in three
groups according to their $[Fe/H]$ and greyscale density distributions
in $\log g, T_{\rm eff}$ and $u-g, g-r$ are produced. Then the three
greyscale pictures are combined to produce two false-color images, one
for the left column and one for the right. The red components in the
false RGB images are based on metal-rich stars with $-0.75 < [Fe/H] <
0$, for the green components intermediate metallicity stars are
selected with $-1.5 < [Fe/H] < 0.75$, finally the most metal-poor
stars with $-3 < [Fe/H] -1.5$ contribute to the blue components of the
images. Therefore, clumps of stars that are mostly blue in the bottom
row of the Figure are mostly metal-poor, the red features are made up
of mostly metal-rich stars, with other colors corresponding to
$[Fe/H]$ mixtures in between. In particular, stars in all three
metallicity bins contributed roughly equal amounts to pixels with dark
grey or almost black color.

The lower left panel of Figure~\ref{fig:tracers} confirms that, as
expected, the BHBs are predominantly metal-poor, the M giants are
metal-rich and the MSTO have no particular metallicity bias. The right
panel in this row showcases beautifully the discriminating power of
the SDSS broad-band filters: the pixels on the stellar locus can be
seen to change their color dramatically depending on their $u-g, g-r$
values. This means that a unique $[Fe/H]$ value can be assigned to a
MS star given its $ugr$ measurements. The idea of photometrically
derived metallicities is the same idea that is behind the UV-excess
method first applied to interpret the chemo-dynamical properties of
the Galactic halo stars by \citet{Eggen1962}. \citet{Ivezic2008}
develop polynomial models (updated recently by \citet{Bond2010}) to
calculate photometric metallicities from SDSS $ugr$ measurements for F
and G Main Sequence stars in the range of $0.2 < g-r < 0.6$.

\subsection{Absolute magnitudes of stellar tracers}
\label{sec:abs_mag}

One of the primary advantages of mapping the galaxy in which we
actually reside is the access to the third spatial dimension. While
most other galaxies appear to us in a cartoonish 2D, distances to the
Milky Way stars can be measured using the annual parallax or with much
cheaper (but less accurate) photometric parallax \citep[see
  e.g.][]{Juric2008}.

Figure~\ref{fig:bhb_msto} reveals exactly how much uncertainty there
exists in determining the luminosities of BHB and MSTO stars using
their broad-band colors only. The left panel of the Figure shows
variation in $g$-band absolute magnitude $M_g$ as a function of $g-r$
color for BHB candidate stars in 11 Galactic star clusters imaged in
the SDSS $ugriz$ filters \citep{An2008}. Regardless of the (small)
metallicity differences and irrespective of the (modest) age spread,
the BHBs form a tight sequence with a gentle slope in $g-r$. The
changes in $M_g$ from slightly above $M_g=0.5$ at red colors to
slightly below $M_g$ at blue colors can be approximated with a simple
polynomial model to give the BHB absolute magnitude within 0.1 mag
\cite[see e.g.][]{Deason2011a}.

The simple $g-r < 0.4$ cut picks up a whole slew of stars of various
luminosities as evidenced by the right panel of
Figure~\ref{fig:bhb_msto}. These range from bright sub-giants at
$M_g\sim 3.5$, through the actual MSTO stars with $3.5 < M_g < 4.5$,
to dwarfs on the Main Sequence that are as faint as $M_g\sim
6$. Additionally, even though the star clusters in the sample
considered do not cover the whole range of metallicity or age,
matching quite well the old and metal-poor Galactic halo, the $[Fe/H]$
and age differences result in significant shifts in both $g-r$ and
$M_g$ around the MS turn-off. As a result, the overall absolute
magnitude spread for the tracers selected is of the order of 3
magnitudes.

It is obvious from the right panel of Figure~\ref{fig:bhb_msto} that
for a star on the Main Sequence, an estimate of the absolute magnitude
can be obtained from the value of its $g-r$ color. The dimming of
dwarf stars with lowering temperature is the basis for the so-called
photometric parallax method, which actually does not have anything in
common with the annual parallactic motion, apart from the fact that it
also delivers the stellar distance. \citet{Juric2008} takes advantage
of the superb quality of the SDSS photometry and calibrates the
absolute magnitude of MS stars using the color-magnitude behavior of
stars in the Galactic globular clusters (GC) with well-determined
distances. Such distance estimate can be further improved, as shown by
\citep{Ivezic2008}, if the photometrically-derived metallicity is
included. However, as shown by \citet{Smith2009, Smith2012}, around
the MSTO the absolute magnitude calibration provided by
\citet{Ivezic2008} suffers from considerable bias. To remedy this,
\citet{Smith2009, Smith2012} offer an appropriately flexible method to
tune the absolute magnitudes of stars around the turn-off according to
their metallicity.

Finally, as regards to M giants, these stars are too luminous, too
metal-rich and too young to be found in the Milky Way's globular
clusters, and hence, the methods of absolute magnitude calibration
discussed above do not apply. However, \citet{Yanny2009} show that
using the pieces of the Sagittarius stream based on the distances
measured with RR Lyrae one can calibrate the M gaint absolute
magnitude to obtain $M_g \sim -1$. Note, however, that this is only
valid for a particular mix of metallicity and age similar to that of
the Sagittarius debris.


\subsection{Matched Filter approach}

Rather than selecting stars in a particular luminosity class and/or
metallicity range to trace the stellar halo sub-structure, an
alternative popular approach is to use the entirety of the stellar
populations belonging to the satellite that is assumed to be
disrupting or disrupted. The probability of any star in the halo to
come from the desired population is obtained by simply taking the
ratio of stellar density in bins of color and magnitude of the
satellite and of the background. These probabilities (or weights) are
then binned on the sky and the smooth slowly-varying component of the
density contributed by the background is subtracted. This so-called
Matched Filter technique as pioneered by \citet{Grillmair1995} has
been employed with great success to isolate extra-tidal congregations
of stars around many Milky Way companions
\citep[e.g.][]{Odenkirchen2001,Rockosi2002,no2010,Sollima2011}. The
method can deliver superb results, but has two inherent breaking
points: i) for satellites that are completely dissolved in the
Galactic gravitational potential, no template color-magnitude density
is available, and ii) as the stars from a disrupting object normally
cover a large area on the sky their heliocentric distances change and
therefore the probabilities assigned by the method will not match
those in the debris everywhere. While the first problem can be easily
overcome by searching for the best-matching CMD template by trial and
error as demonstrated beautifully by \citet{Grillmair2006a}, there is
no simple (and elegant) remedy to the issue of evolving distance.

\section{Stellar halo of the Galaxy}

\subsection{Evidence of sub-structure in the stellar halo}

\subsubsection{The Field of Streams}

\begin{figure}
\centering
\includegraphics[width=0.99\linewidth]{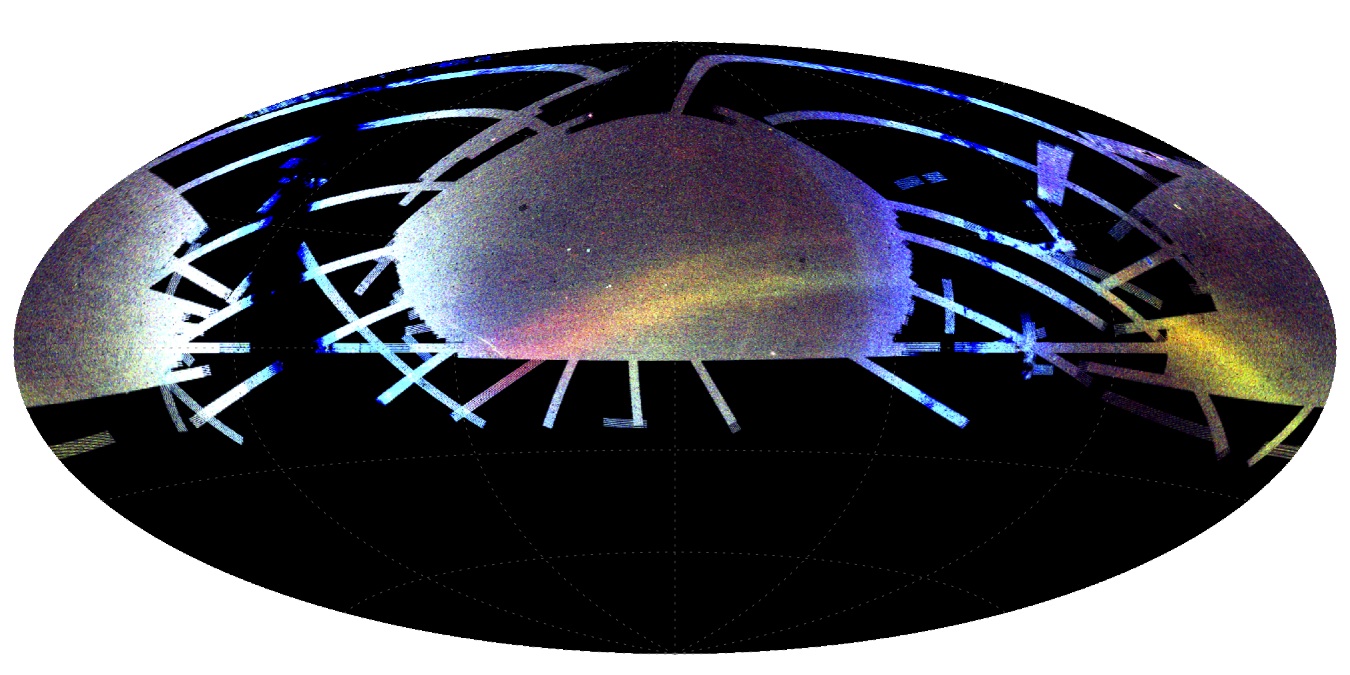}
\includegraphics[width=0.32\linewidth]{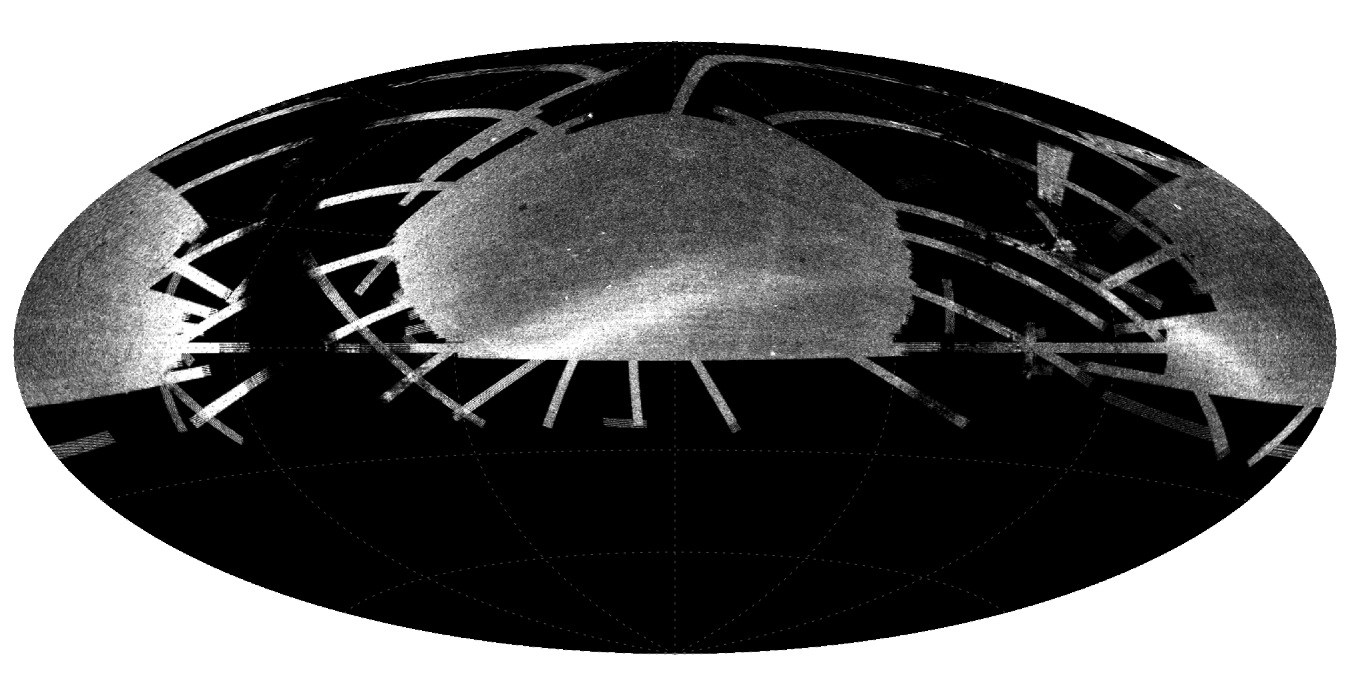}
\includegraphics[width=0.32\linewidth]{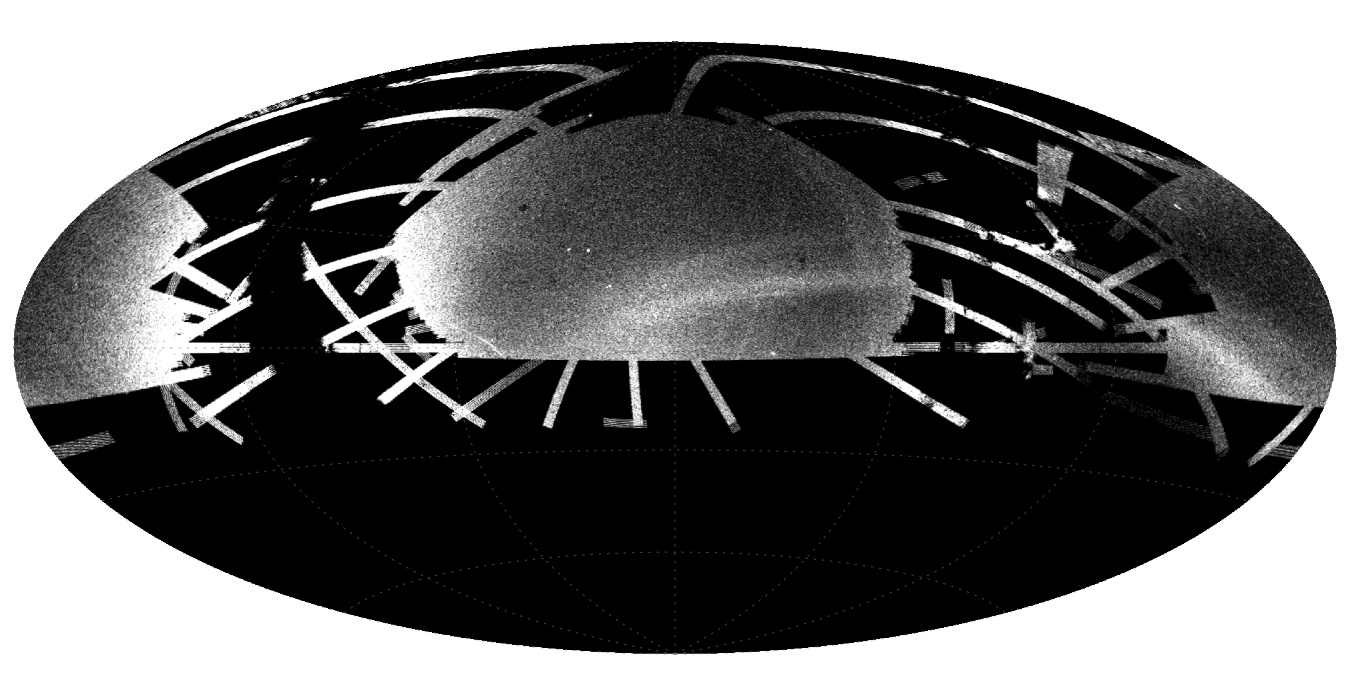}
\includegraphics[width=0.32\linewidth]{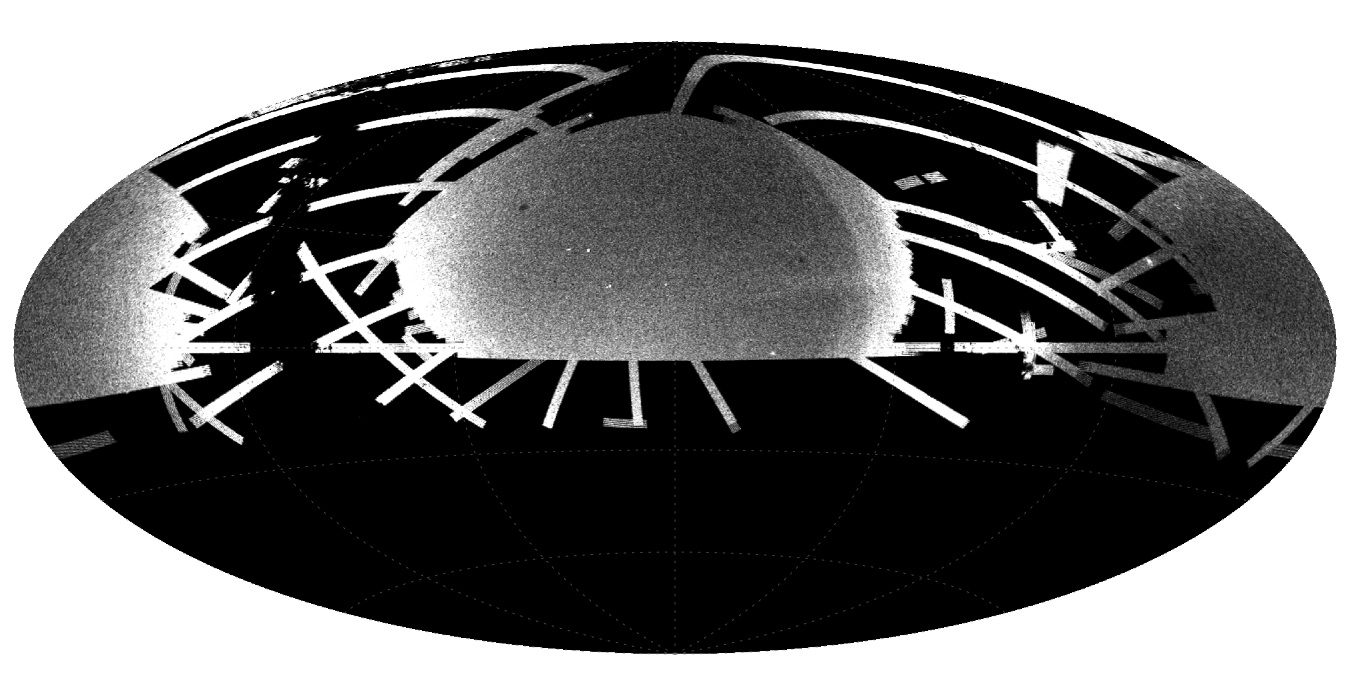}
\caption{Map of the stellar halo of the Milky Way in Equatorial
  coordinates (centered on RA$=180^{\circ}$) using the data from SDSS
  DR9. {\it Top:} False-color composite image uses $\sim$ 16,000,000
  predominantly MS and MSTO stars selected with a simple color cut
  $g-i<0.6$ and split into three equal magnitude bins between $i=19$
  and $i=22.5$. Given the $g-i$ cut, the color of the pixel can be
  interpreted as a rough estimate of the average heliocentric distance
  of the stars contributing to it. White corresponds to high stellar
  densities in all three magnitude bins. Black corresponds to areas
  with missing data.{\it Bottom:} Grey-scale images each showing the
  stellar tracer density in a particular magnitude bin are then
  combined in the order that maps the densities of bright,
  intermediate and faint stars onto blue, green and red channels as
  shown in the Top panel. }
\label{fig:fos_equ}
\end{figure}

\begin{figure}
\centering
\includegraphics[width=0.99\linewidth]{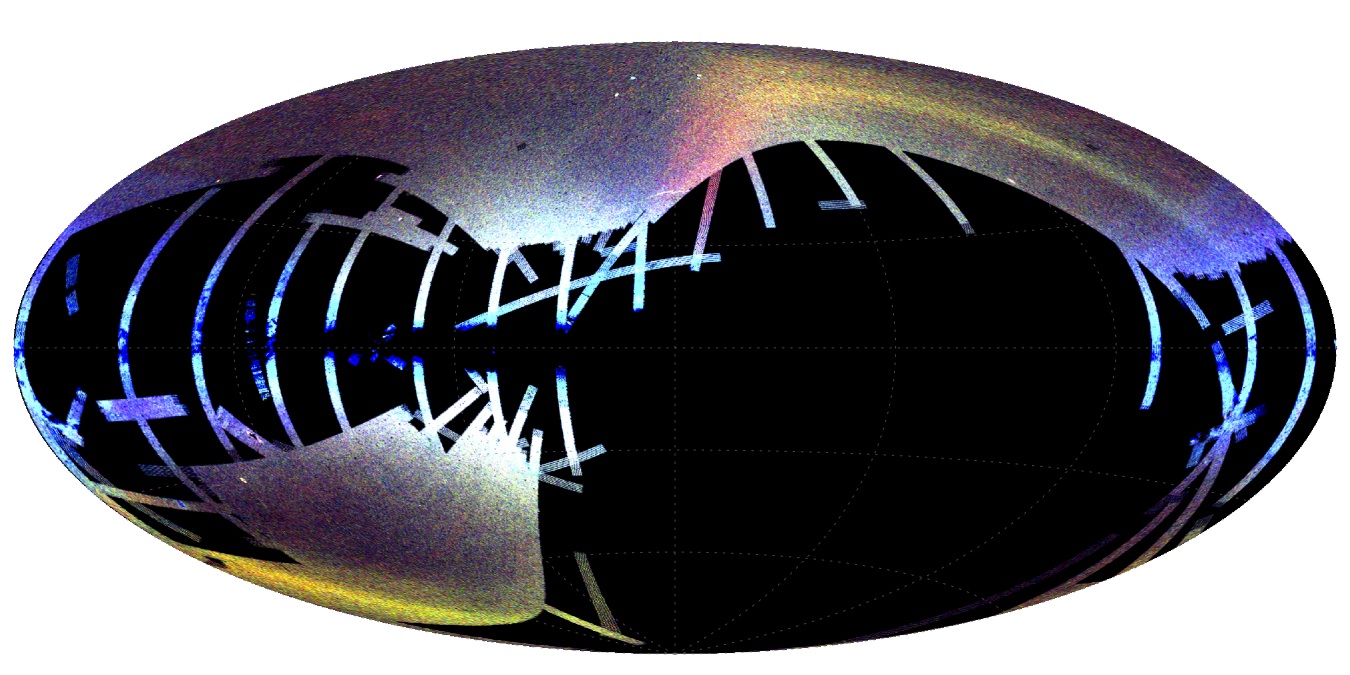}
\includegraphics[width=0.32\linewidth]{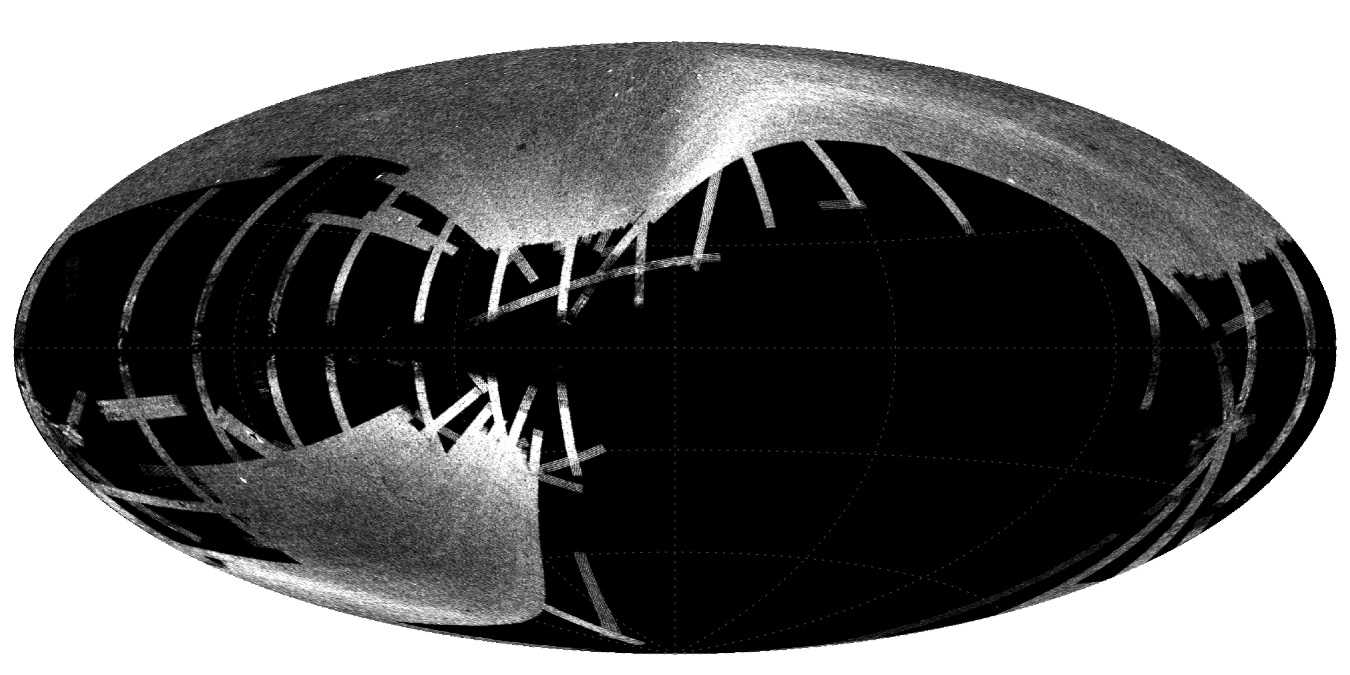}
\includegraphics[width=0.32\linewidth]{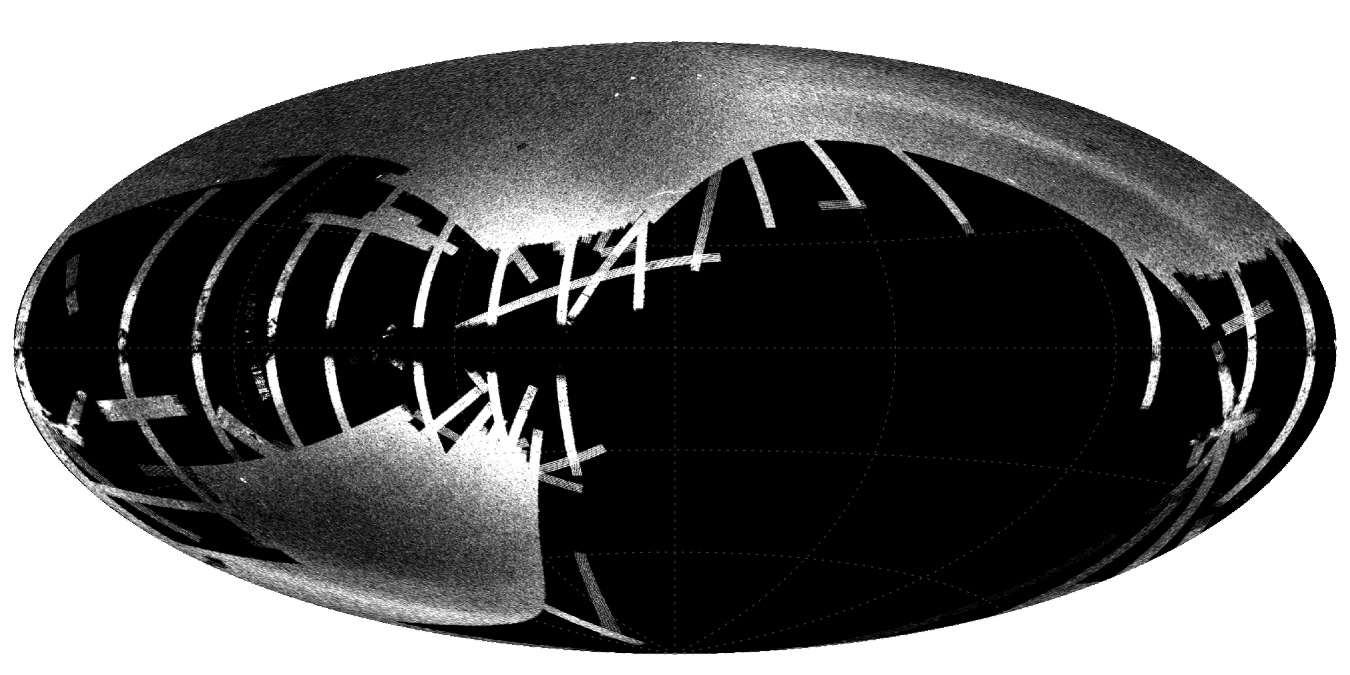}
\includegraphics[width=0.32\linewidth]{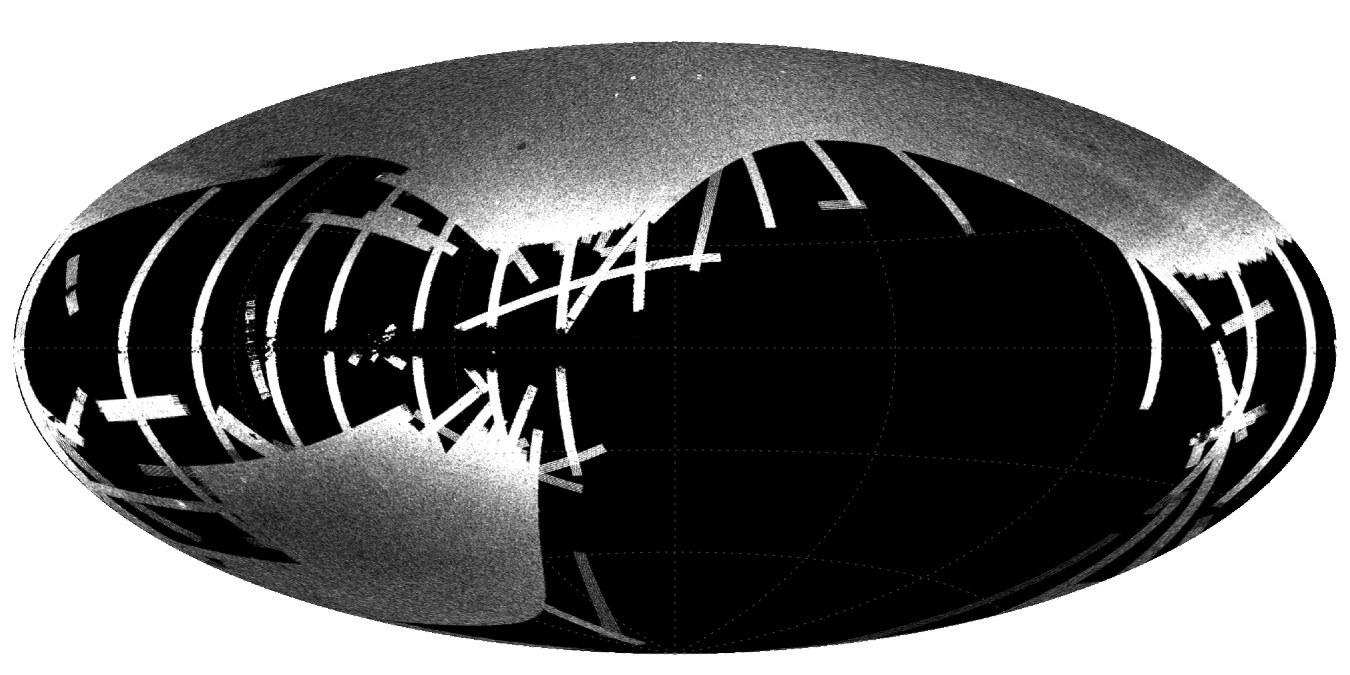}
\caption{Same as Figure~\ref{fig:fos_gal} but in Galactic
  coordinates. Galactic $l=0^{\circ}, b=0^{\circ}$ is at the centre of
  the Figure.}
\label{fig:fos_gal}
\end{figure}

The broad-brush spatial properties of the sub-structure in the Milky
Way's halo as seen by the SDSS in its DR9 incarnation \citep{Ahn2012}
are displayed in Figures~\ref{fig:fos_equ} and ~\ref{fig:fos_gal} in
Equatorial and Galactic coordinates respectively. The SDSS DR5 version
of this false-color halo map dubbed the Field of Streams is published
by \citet{Belokurov2006a}. Included in this map are predominantly MS
and MSTO stars, which are selected with a simple color cut $g-i<0.6$
(a selection almost identical to the described above $g-r<0.4$). These
stars are then split into three equal magnitude bins between $i=19$
and $i=22.5$. Three grey-scale images each showing stellar tracer
density in a particular magnitude bin (see bottom row of the Figures)
are then combined in the order that maps bright-intermediate-faint
stars onto blue-green-red channels. Given the fixed $g-i$ cut, the
false RGB color of each pixel can be interpreted as a rough estimate
of the average heliocentric distance of the stars contributing to it.
Assuming, very approximately, that the typical $M_i \sim 4.$ for the
selected tracers, given the magnitude range of $19 < i < 22.5$, the
range of the heliocentric distances probed is $10< D {\rm (kpc)} <
50$.

Most of the contiguous sky coverage in the SDSS footprint falls around
the North Galactic Cap. The arc dominating the density map at the high
Galactic latitudes in the North is the leading tidal tail emanating
from the Sagittarius dwarf galaxy. The Sgr debris is also the only
prominent structure in the stellar halo in the Galactic South, where
the trailing tail of the dwarf can be seen. In
Figure~\ref{fig:fos_equ}, the leading stream changes color from red to
green-blue as its heliocentric distance drops from $\sim 50$ kpc at
Dec$=220^{\circ}$ (just behind the Galactic centre) to $\sim 15$ kpc
at Dec$=130^{\circ}$ \citep{Belokurov2006a}. There, at the Galactic
anti-centre, the Sgr stream crosses the prominent Galactic Anti-Center
Stellar Structure seen in the Figure as a violet-blue tilted band with
striation. This complex configuration is due to the multiple
components of the GASS: the so-called Monoceros Ring
\citep[e.g.][]{Newberg2002}, the Anti-Center Stream
\citep[e.g.][]{Grillmair2006d, Grillmair2008} and the Eastern Banded
Structure \citep{Grillmair2006d}. The fuzzy green haze directly
underneath the Sgr stream at around RA$\sim 180^{\circ}$ is the large
cloud of stars dubbed Virgo overdensity \citep[e.g.][]{Juric2008}. The
central, dense regions of the stellar halo can be seen as bright
white-blue glow on either size of the Galactic disk at $320^{\circ} <$
RA $< 220^{\circ}$. As \citet{Belokurov2007a} point out, the
distribution of the MS and the MSTO stars does not peak in the
direction of the Galactic centre as it seems offset towards the
positive Galactic $l$. Moreover, on closer examination there appears
to be a substantial asymmetry in the counts of MS/MSTO stars with
heliocentric distances $10 < D {\rm (kpc)}< 20$ between the Galactic
North and the South. These observations lead the authors to the
conclusion that a sizable portion of the central Milky Way halo could
be due to yet another massive stellar sub-structure, so-called
Hercules-Aquila Cloud.

A Galactic projection of the same map (see Figure~\ref{fig:fos_gal})
helps to make sense of some of the stellar halo density patterns. For
example, it shows that the Sagittarius leading tail nearly misses the
Galactic North, the GASS is mostly confined to Galactic latitude $b <
30^{\circ}$. Additionally, even though the SDSS coverage at positive
and negative Galactic longitude is not equal, it is clear that at
$l>0^{\circ}$ at high $b>30^{\circ}$ there does not exist a
counter-part to the Virgo overdensity.

\subsubsection{The big 4}
\label{sec_big4}

The Sagittarius Stream, the Galactic Anti-center Stellar Structure,
the Virgo and the Hercules-Aquila Clouds are the four largest stellar
structures in the halo of the Milky Way. Out of these four, only the
Sgr Stream lies predominantly outside the Galactic disk making it
possible to estimate its total extent and the overall stellar
mass. The Stream consists of two tails, the leading and the trailing,
flowing from the Sgr dwarf galaxy, which currently lies on the
opposite side of the Galaxy, behind the bulge, several degrees under
the disk. The dwarf is falling onto the disk and has just passed its
point of the nearest approach at $\sim$15 kpc from the Galactic
center. The two tails appear bifurcated \citep[see
  e.g.][]{Belokurov2006a,Koposov2012} and extend each at least as far
as $\sim$ 180$^{\circ}$ away from the progenitor (see
Figures~\ref{fig:fos_equ} and \ref{fig:fos_gal}). The leading tail is
traced as far as 50 kpc from the Galactic center, while the
apo-galacticon of the trailing debris is probably as far as 60-100
kpc. Both the Sgr remnant and the stream host a range of stellar
populations with different ages and metallicities. In particular,
along the stream, a substantial population gradient is observed
\citep[e.g.][]{Chou2007,Yanny2009,Bell2010,Chou2010,Keller2010,Carlin2012},
which, within any sensible model of the dwarf disruption, would mean a
similarly pronounced abundance and age gradient in the
progenitor. Using a variety of stellar tracers across the sky,
\citet{No2010b} map the Sgr debris and, correcting for the distance
and the abundance gradients estimate the total stellar luminosity of
the progenitor prior to disruption. They find that, before merging
with the Galaxy, the dwarf was as bright as $1.5\times 10^8 M_{\odot}$
or just under $M_V \sim -16$, but today it has lost as much as $70\%$
of its stars to the Galactic tides.

The Virgo Cloud can be seen as green haze directly underneath the Sgr
Stream at around $RA\sim 12^h$. While early glimpses of this structure
are reported in several studies, based on the SDSS DR4 imaging data,
\citet{Juric2008} provide the first large scale map of the Cloud and
emphasize its gigantic extent on the sky of least $\sim
1000$ deg$^2$. From the inspection of Figure~\ref{fig:fos_equ}, it is
obvious that the portion of the Virgo Cloud analyzed by
\citet{Juric2008} is only the tip of the structure that appears to
continue to lower Declinations as far as the SDSS/Segue imaging
stripes can go. Accordingly, \citet{Bonaca2012b} take advantage of the
extra imaging in the SDSS DR8 and claim that the true extent of the
Cloud is somewhere between 2000 deg$^2$ and 3000 deg$^2$. The
debris cover an enormous portion of the sky, but given the typical
distance and the low surface brightness, the total luminosity of the
Virgo Cloud is estimated to be modest $< 10^6 M_{\odot}$
\citep{Bonaca2012b}.

The Galactic Anti-Center Stellar Structure and the Hercules-Aquila
Cloud have most of their stars at low Galactic latitudes: within $|b|
< 40^{\circ}$, GASS can be found at roughly $120^{\circ} < l <
240^{\circ}$ and HAC at $20^{\circ} < l < 70^{\circ}$ (see
Figure~\ref{fig:fos_gal}. In fact, both of these structures appear to
be stuck right in the plane of the disk as their candidate member
stars are detected in both Northern and Southern hemispheres. Given
such an awkward location in the Galaxy, it is still questioned whether
all, or at least some of the signal attributed to these two can be
explained away invoking variants of the known components of Milky
Way. For example, it is claimed that parts of the GASS can well be
ascribed to the Galactic flare and/or the warp
\citep[e.g.][]{Ibata2003}, and the HAC is really nothing but the
asymmetric thick disk \citep[e.g.][]{Larsen2008,Larsen2011}. However,
there exists additional observational data within which stellar
over-densities are clearly seen in the directions of both the GASS and
the HAC in tracers unlikely to populate either of the disks. For
example, the distant portion of the GASS, the And-Tri stream is traced
with M giants at distances of the order of 30 kpc. HAC can be picked
up with RR Lyrae in the SDSS Stripe 82 dataset
\citep[e.g.][]{Watkins2009, Sesar2010a} at $10 < D < 20$ kpc.  As most
of the light in both GASS and HAC is hidden in the Galactic plane,
only very approximate estimates of their total stellar masses exist in
the literature. \citet{Belokurov2007b} give a conservative estimate of
$\sim 10^7 L_{\odot}$ for the Hercules-Aquila Cloud. For the closer
portion of the GASS, \citet{Yanny2003} get the total stellar mass in
the range of $0.2 - 5 \times 10^8 M_{\odot}$, with the larger value
obtained assuming that i) the GASS follows an exponential profile as a
function of z and ii) encompasses the entire Milky Way. Several
follow-up studies present the updated measurements of the structure
and the stellar populations of the pieces of GASS visible in the SDSS
\citep[e.g.][]{Dejong2010,Grillmair2011,Li2012} and in the deeper
imaging \citep[e.g.][]{Conn2012}. According to the body of work
published so far, the components of the GASS most consistent with the
accretion scenario \citep[see e.g.][]{Penarrubia2005} have, overall,
much flatter density distribution as a function of Galactic $|b|$ or
$|z|$. If true, this observation would lead to the substantial
reduction of the overall luminosity of the GASS. Perhaps, the
following simple argument can be constructed to provide a
complementary guess as to the total stellar mass in the GASS. Given
that the parts of the GASS detected within the SDSS field of view
typically have similar or lower surface brightness as compared to the
Sgr Stream, but are on average closer by a factor of 2-5, it is not
unlikely that the structure, in fact, contains more than $10^8
M_{\odot}$.

\subsubsection{Ultra-faint satellites}

\begin{figure}
\centering
\includegraphics[width=0.99\linewidth]{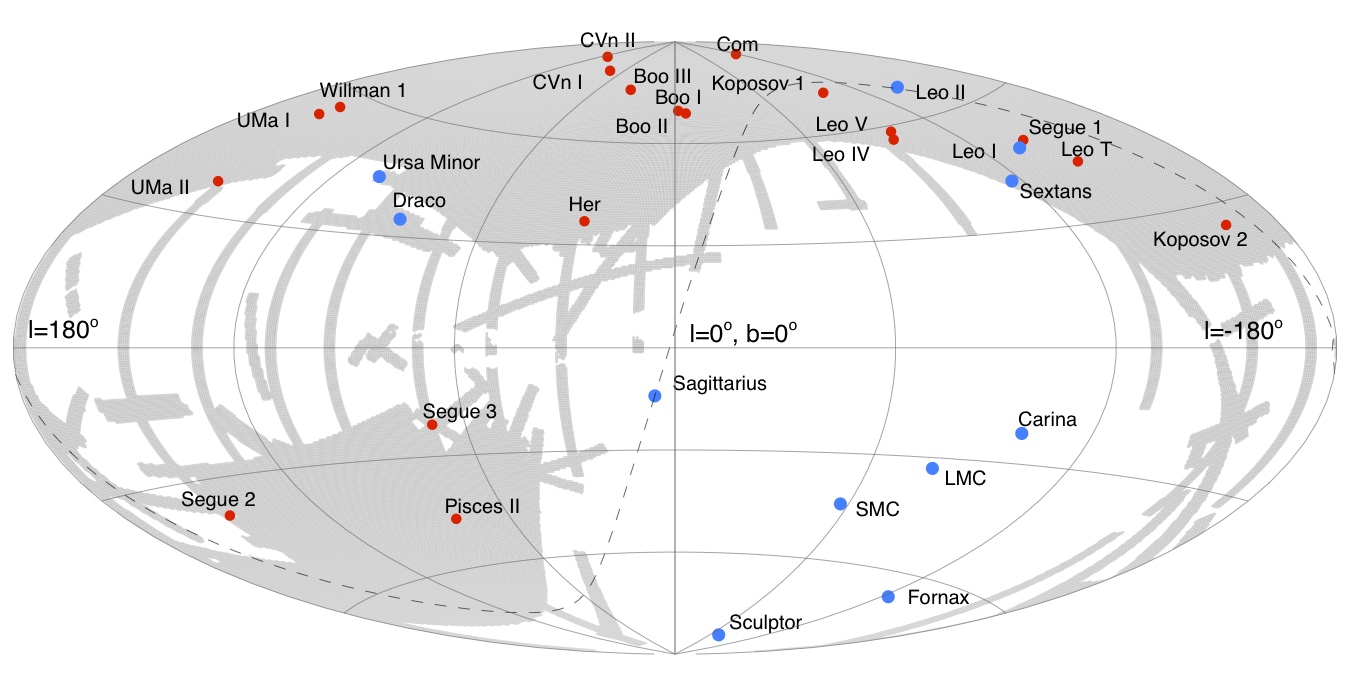}
\caption{Distribution of the classical dwarf galaxies (blue filled
  circles) and the SDSS ultra-faint satellites (red filled circles),
  including three ultra-faint star clusters, in Galactic
  coordinates. The SDSS DR8 imaging footprint is shown in grey. Dashed
  line marks the tentative orbit of the Sgr dwarf galaxy. Galactic
  $l=0^{\circ}, b=0^{\circ}$ is at the centre of the Figure.}
\label{fig:dwarfs_lb}
\end{figure}

Visible as bright dots of different colors in the maps in
Figures~\ref{fig:fos_equ} and~\ref{fig:fos_gal} are the compact
stellar over-densities corresponding to the Galactic satellites that
give the impression of being still intact. The brightest of these
``hot pixels'' correspond to the well-known star clusters and
classical dwarf galaxies, while the very faint and barely visible
small-scale over-densities mark the locations of the so-called
ultra-faint satellites of the Milky Way. Although several of these,
including Boo I, Boo III, CVn I and UMa II, are seen in this picture
with a naked eye, the rest of the population of these objects is too
insignificant and can only be unearthed via an automated over-density
search. The first example of such an automated stellar over-density
detection procedure is presented in \citet{Irwin1994} who apply the
method to the data from the photographic plates of the POSS I/II and
UKST surveys scanned at the APM facility in Cambridge. A vast area of
20,000 square degree of the sky is searched but only one new nearby
dwarf galaxy is detected, namely the Sextans dSph. A variant of the
procedure is used, albeit with a little less luck, by
\citet{Kleyna1997}, and subsequently by \citet{Willman2005a,
  Willman2005b} who actually find the two very first examples of
ultra-faint objects in the SDSS data. The ease with which these
systems reveal themselves in a stellar halo density map akin to the
``Field of Streams'' \citep[see][]{Zucker2006a, Belokurov2006c} helped
to re-animate the search for new Milky Way satellites and more than a
dozen of new discoveries have been reported in quick succession
\citep{Zucker2006b,Belokurov2007c,Irwin2007,Koposov2007,Walsh2007,
  Belokurov2008,
  Belokurov2009,Grillmair2009,Belokurov2010}. Figure~\ref{fig:dwarfs_lb}
maps the distribution of all presently known SDSS ultra-faint
satellites on the Galactic sky.

The accuracy and the stability of the SDSS photometry makes it
possible for the over-density detection algorithms to reach
exceptionally faint levels of surface brightness across gigantic areas
of the sky. However, even though genuine Galactic satellites can be
identified in the SDSS as groups of only few tens of stars, their
structural parameters can not be established with adequate accuracy
using the same data. Deep follow-up imaging on telescopes like INT,
CFHT, LBT, Magellan, MMT, Subaru and most recently HST, has played a
vital role in confirming the nature of the tiny stellar blobs in the
SDSS, as well as in pinning down their precise sizes, ellipticities
and their stellar content. The most recent, deep and wide photometric
studies of a significant fraction of the new SDSS satellites are
published by \citet{Okamoto2012} and \citet{Sand2012}. They point out
that even at distances $D>100$ kpc from the Galactic centre, the outer
density contours of CVn II, Leo IV and Leo V display extensions and
perturbations that are probably due to the influence of the Milky Way
tides. Similarly, there is now little doubt that both UMa II and Her
are excessively stretched, as their high ellipticities as first
glimpsed at discovery \citep{Zucker2006a, Belokurov2007c} are
confirmed with deeper data \citep{Munoz2010, Sand2009}. Note, however
that apart from these two obvious outliers there does not seem to be
any significant difference in the ellipticity distributions of the
UFDs and the Classical dwarfs contrary to the early claims of
\citet{Martin2008}. This is convincingly demonstrated by
\citet{Sand2012} with the help of the imaging data at least 2
magnitudes deeper than the original SDSS.  They, however, detect a
more subtle sign of the tidal harassment: the preference of the
density contours of the SDSS satellites to align with the direction to
the Galactic centre.

As far as the current data is concerned, the SDSS dwarfs do not appear
to form a distinct class of their own, but rather are the extension of
the population of the Classical dwarfs to extremely faint absolute
magnitudes. However, as more and more meager luminosities are reached,
it becomes clear how extreme the faintest of the UFDs are. The
brightest of the group, CVn I and Leo T show the usual for their
Classical counter-parts signs of the prolonged star-formation. For
example, CVn I hosts both Blue Horizontal Branch and Red Horizontal
Branch populations, while Leo T shows off a sprinkle of Blue Loop
stars. However, the rest of the ensemble appears to have narrow CMD
sequences with no measurable color spread around the conventional
diagnostic features, e.g. MSTO and/or RGB, thus providing zero
evidence for stellar populations born at different epochs
\citep[e.g.][]{Okamoto2012}. The CMDs of the UFDs have revealed no
secrets even under the piercing gaze of the HST: all three objects
studied by \citet{Brown2012} appear to be as old as the ancient
Galactic globular cluster M92. Yet the low/medium and high-resolution
follow-up spectroscopy reveals a rich variety of chemical abundances
somewhat unexpected for such a no-frills CMD structure. The first
low-resolution studies of \citet{Simon2007} and \citet{Kirby2008}
already evince the existence of appreciable $[Fe/H]$ spreads in the
SDSS dwarfs with the metallicity distribution stretching to extremely
low values. Analyzing the medium and high resolution spectra of the
Boo I system, \citet{Norris2010} measure the spread in $[Fe/H]$ of
$\sim$1.7 and the $[Fe/H]$ dispersion of $\sim$0.4 around the mean
value of -2.55 at $M_V\sim -6$. It seems that this behavior of
decreasing mean metallicity with luminosity while maintaining a
significant enrichment spread is representative of the UFD sample as a
whole \citep[see also][]{Lai2011,Koch2013,Vargas2013}. Crucially,
these spectroscopic observations require that, notwithstanding their
low stellar luminosities at the present day, these satellites had
enough total mass in the past to hold on to some of the enriched gas
after the first supernovae explosions and subsequently produce more
stars. Additionally, in the UFDs, the downwards shift of the mean
metallicity with decreasing stellar mass reveals that they can not
simply be direct analogs of the Classical dwarfs stripped off the bulk
of their stellar content.

Of the 16 ultra-faint satellites currently known, only 5 systems have
a handful of stars studied with high-resolution spectroscopy. More
specifically, one star in Leo IV \citep{Simon2010}, two stars in Her
\citep{Koch2008}, 3 stars in each of UMa II and Com \citep{Frebel2010}
and 7 in Boo I \citep{Gilmore2013} have been measured so far. It is
perhaps too early to draw far-reaching conclusions from these highly
incomplete data, nonetheless an interesting picture seems to be
emerging from the detailed abundance work. Although wanting in
quantity, these high-resolution high-quality spectroscopic data do
robustly confirm the key properties of the UFD chemical enrichment
histories hinted at by the analysis of the low-resolution (and at
times, low-S/N) samples. The SDSS dwarfs are indeed characterized by
remarkably low levels of the overall iron enhancement as well as the
heterogeneity of the individual stellar abundances (in each of the 4
satellites that have more than 1 star measured). Additionally, the
very first high-resolution study of a UFD by \citet{Koch2008} reported
a depletion of heavy neutron capture elements. RGB stars with low
abundance levels of barium are also found in Leo IV, Com, UMa II and
Boo I \citep{Simon2010, Frebel2010, Gilmore2013}. Moreover, in Boo I,
several extremely metal-poor stars are demonstrated to have increased
levels of carbon \citep[see e.g.][]{Norris2010}. Potentially, there
are at least two notable implications of these enrichment
patters. First, carbon-enhanced metal-poor stars are common denizens
of the Galactic stellar halo, yet if there occur any in the classical
dSphs, they have so far eluded the detection. The existence of such
stars in both the UFDs and the MW stellar halo may signify the
commonality of the chemical evolution paths of the halo progenitor(s)
and the ultra-faint satellites. Second, as several authors have
pointed out \citep[e.g.][]{Koch2008,Simon2010, Frebel2010}, the
enhancement in $\alpha$-elements together with the depletion in
neutron-capture elements at low metallicities can be linked to the
products of the Population III SNe, therefore implying that a good
fraction of the stellar content in the UFDs could be direct
descendants of the first stars \citep[see also][]{Frebel2012}.

It is evidently not possible to come up with a sensible theory of the
UFD formation without an idea of their total masses. Such a
measurement, which necessarily involves accurate kinematics for a
large enough sample of the satellite members, is, however, not
straightforward. This is simply due to the fact that, as illustrated
by \citet{Koposov2008}, the majority of these objects are discovered
very close to the detection boundary, implying that the over-density
signal is dominated by the stars close to the SDSS detection limit of
$r\sim 22$. At these magnitudes, only half a handful of facilities in
the world are capable of obtaining absorption spectra of
signal-to-noise sufficient to measure the line-of-sight velocities of
individual stars. Even if the kinematic signal is present in the data,
winnowing it out from the low-resolution spectra of low-metallicity
stars is a challenge. An even harder challenge is figuring out the
uncertainties of the velocity measurements. For most ultra-faints, the
typical member velocity uncertainty is of the order of, or larger
than, the intrinsic velocity dispersion of the system. Under or
over-estimating the measurement error by a small fraction can lead to
a substantial systematic velocity dispersion bias, and as a
consequence, a wrong aperture mass. Despite the above mentioned
difficulties of the task at hand, several teams report the results of
their heroic attempts to gauge the central masses of the UFDs
\citep[e.g.][]{Martin2007b, Simon2007, Walker2009, Belokurov2009,
  Simon2011, Koposov2011}

The structural parameters of the faintest of the SDSS satellites,
e.g. Willman 1, Segue 1 and 2, Boo II are dangerously similar to those
of the most diffuse star clusters in the Milky Way and M31. It is not
conceivable, purely on the basis of their size or luminosity, to come
up with the most likely scenario of their formation. Therefore, their
kinematic and chemical properties are the most important clue. Today,
for the faintest objects, it is just possible, after many hours spent
on Keck and VLT, to build datasets with radial velocities for a dozen
or two of the MSTO members and a trickle of the Red
Giants. Accordingly, the most recent and the most thorough kinematic
analysis of Willman 1, Segue 1 and Segue 2 can be found in
\citet{Willman2011, Simon2011} and \citet{Kirby2013}
correspondingly. Moreover, \citet{Norris2010} independently carries
out a thorough chemical study of Segue 1 using a different combination
of the telescope, the instrument and the analysis techniques. For
these three best studies objects, the picture does not appear to be as
clear-cut as for their more luminous peers. For example, the evolution
of the line-of-sight velocity with radius in Willman 1, where the
inner-most stars are offset by some 8 km/s from the outer-most ones is
unusual, and is, perhaps, a sign of the advanced stage of tidal
disruption. There is also an evidence of the spread in [Fe/H], but
unfortunately it is based on the measurements of only two Red Giant
Branch stars.

Segue 1, the best studied of the three, has a substantial velocity
dispersion at 3.7$^{+1.4}_{-1.1}$ km/s and an impressive metallicity
spread. There are however some quirks with regards to both the
velocity and the metallicity dispersion measurements, such as the fact
that the velocity dispersion calculated using the brightest members
only (5 red giants stars) is essentially consistent with zero, or the
fact that some of the most metal-poor stars also lie several
half-light radii away from Segue 1's center
\citep[see][]{Norris2010}. Perhaps more significant is the observation
by \citet{Newberg2010} that the Orphan stellar stream runs at the
identical distance and velocity only $\sim$2 degrees away from the
position of Segue 1. Given the width of the stream of 1 degree, a
significant contamination of spectroscopic samples at Segue 1's
location is not very likely. Yet, the dynamical association between
the two is, however, quite possible: both the progenitor of the Orphan
Stream and Segue 1 itself might have been parts of a bigger system
which is now completely disrupted.

The evidence of such an accretion event is even more dramatic in the
case of Segue 2. Taking into account the observations reported in
\citet{Majewski2004,Rocha2004}, Segue 2 is immersed in the debris of
the Triangulum-Andromeda stream, which is interpreted as the distant
(at $\sim$ 30 kpc compared to $34$ kpc for Segue 2) counter-part of
the Monoceros stream and part of the larger Galactic Anti-Center
Stellar Structure. As published by \citet{Rocha2004}, the velocities
of M giant members of Tri-And structure are $0 < V_{GSR} < 60$ in the
range of longitudes $160^{\circ} < l < 130^{\circ}$ at the Galactic
latitudes slightly lower than that of Segue 2. This velocity
distribution can be modeled as a Gaussian that peaks around $V_{GSR}
\sim 30$ km s$^{-1}$ which is a good match to the measurement of the
satellites line-of-sight velocity $V_{GSR} \sim 40$ km s$^{-1}$. The
coverage of the area with the spectroscopic M giants is sparse, and
the SDSS spectroscopic footprint is seriously incomplete
here. However, \citet{Belokurov2009} present an unambiguous kinematic
evidence for the stream's existence using the spectra obtained with 1
degree wide field Hectochelle instrument on MMT. They claim that the
stream's stars are more metal-rich on average and their velocity
distribution can be described with a broader Gaussian, namely 15 km/s
vs $\sim$3 km/s for Segue 2. Most recently, \citet{Kirby2013}
re-evaluated the spectroscopic properties of Segue 2 albeit with a
different observational setup and a smaller field of view as compared
to the original study of \citet{Belokurov2010}. They claim no
detection of the stream signal, which is perhaps not surprising given
the targeting strategy and the area of the sky surveyed. Intriguingly,
they measure much lower velocity dispersion (essentially consistent
with zero), thus markedly reducing the central mass of the satellite.

\subsubsection{Star cluster streams}

The large undissolved stellar clouds (Virgo, Hercules-Aquila) and
broad long streams (Monoceros, Sagittarius) described earlier are the
primary contributors to the Galactic halo in terms of the stellar
mass. In the past decade, an assortment of much narrower, often
shorter and significantly less luminous streams has been
identified. It seems most likely that these would have originated in
star clusters. Some of these wispy tidal tails are discernible in
Figures~\ref{fig:fos_equ} and~\ref{fig:fos_gal}, such as the tidal
tails of the Palomar 5 globular cluster
\citep{Odenkirchen2001,Grillmair2006b}. However, in their majority
these feathery streams require a more subtle approach and are best
seen with the help of the Matched Filter technique. Some of the star
cluster debris have obvious progenitors like the short stubby tails
visible around e.g. NGC 5466 \citep{Belokurov2006b}, NGC 5053
\citet{Lauchner2006}, Pal 14 \citep{Sollima2011}, Pal 1
\citep{no2010}. For the others, typically extending many degrees on
the sky, no suitable progenitor has been discovered yet, e.g. the GD-1
stream \citep{Grillmair2006a}, a group of four streams Styx, Acheron,
Cocytos, Lethe \citep{Grillmair2009} and the most recently identified
Pisces Stellar Stream \citep{Bonaca2012, Martin2013}.

It is interesting to estimate the total number of star clusters that
have disrupted so far and whose stars are now part of the Galactic
halo. While such a count is valuable as it gives an idea of the
fraction of the halo that is comprised of the GC debris, it is not
straightforward as it requires the knowledge of the Cluster Initial
Mass Function (CIMF) and a model of the cluster evolution in the Milky
Way tidal field. An example of such calculation is presented in
\citet{Poul2013} who approximate the CIMF with a power-law
distribution and apply the semi-analytic model of \citet{Gieles2011}
for the star cluster evolution in a logarithmic Galactic
potential. They find that, of the several models they consider, the
Roche volume under-filling model with a flat CIMF (power law index 0)
reproduces the present day properties of the Milky Way's GCs the
best. While the authors do not give the exact number of dissolved
clusters, it is clear that the flat mass function evolution can only
produce a moderate number of star cluster streams in the Galactic
halo, perhaps orders of magnitude less as compared to the rising power
laws (e.g. -2). Alternatively, the number of the GC streams detected
so far with the SDSS could be translated into a Galaxy total if there
existed an estimate of the stream detection efficiency. However, it is
possible that a significant fraction of the known long and narrow
stellar streams may have been produced as a result of only a few
accretion events. For example, given the noticeable alignment of their
orbital planes, it is feasible that the progenitors of the Styx,
Acheron, Cocytos and Lethe streams arrived to the Galaxy together with
a much bigger satellite. The fact that the GC accretion is most likely
linked to the infall of more massive Galactic satellites is another
reason to believe that the total number of GC streams is relatively
low given the evidence for the uneventful Milky Way's mass assembly
history.

\subsubsection{Orphan and Styx. Streams from ultra-faint satellites?}

The tidal stream's cross-section on the sky is normally a giveaway of
the progenitor's mass. The low-density disrupting star clusters with
small internal velocity dispersion $\sigma \lesssim 1$ km s$^{-1}$
typically produce tails that are only $\sim 0.1^{\circ}$ wide. On the
other hand, a galaxy as massive as Sgr dwarf with its current $\sigma
\lesssim 20$ km s$^{-1}$ \citep[see e.g.][]{Ibata2009} gives rise to
streams that are at least 10$^{\circ}$ across (see
Figure~\ref{fig:fos_equ} for example). This rule of thumb of course
assumes comparable distances to the tidal tails and not hugely
different dynamical ages. Depending on how aspherical the
gravitational potential of the Galaxy is and how long ago the debris
were stripped, even originally narrow tails can puff up with time.

Amongst the panoply of stellar substructure recently discovered in the
Galactic halo, there are at least two streams that seem to occupy the
parameter space intermediate between the star clusters and dwarf
galaxies. These are the Orphan stream \citep{Belokurov2006a,
  Belokurov2007b, Grillmair2006c} visible in Figure~\ref{fig:fos_equ}
as almost vertical streak of orange color crossing the Sgr debris at
around $140^{\circ} <$ RA $< 160^{\circ}$, and the Styx stream
\citep{Grillmair2009}, the faint blue nebulous smear running at almost
constant Dec$=30^{\circ}$ from RA$=250^{\circ}$ to RA$=220^{\circ}$
where it starts to drop in Dec towards the Sgr stream. Curiously, both
Orphan and Styx run in a close vicinity of the several of the Galactic
ultra-faint satellites. The sky projection of the orbit of the Orphan
stream takes it right through the position of the UMa II dwarf. The
feasibility of such association is explored in \citet{Fellhauer2007}
who conclude that UMa II could well be the stream's
progenitor. However, as convincingly shown in \citet{Newberg2010}, the
early tentative estimates of the stream's radial velocity were
incorrect and that the actual orbit of the stream is much more
consistent with the 4D location of Segue 1. As regards to the Styx
stream, when tracing its debris to the lower RA, \citet{Grillmair2009}
discovers a pronounced stellar clump within the stream's path. Dubbed
Bootes III and subsequently confirmed with spectroscopy
\citep{Carlin2009} this is the most diffuse of all ultra-faints found
so far.

\subsubsection{Broad and Invisible}

As the proper motion, spectroscopy and the variability wide-area
surveys slowly catch up with the rapidly advancing sky imaging
campaigns, it is possible to gauge the presence of stealth stellar
structures, so diffuse that they elude detection in stellar halo maps
akin to those described above. These detections are reminiscent of the
original discovery of the Sgr dwarf \citep{Ibata1994} that is too
faint and spread out to be seen on a photographic plate but produces a
booming signal in radial velocities.

Trinagulum-Andromeda is an extended stellar structure located at
several tens of kpc from the Galactic centre \citep{Rocha2004}. It is
initially picked up as a faint excess of 2MASS M-giant stars, and
later confirmed with the help of proper motion data and follow-up
spectroscopy. As judged by the radial velocities of its members, the
Tri-And cloud seems to be connected to the Southern Galactic
counterpart of the Monoceros stream, and thus forms the more distant
wraps of the Galactic Anti-centre Stellar Structure
\citep{Newberg2002, Ibata2003, Rocha2003,
  Yanny2003}. \citet{Majewski2004} and \citet{Martin2007} report the
detection of the Main Sequence stars in the Tri-And cloud, thus
ridding of the last shreds of doubts as to the reality of its
existence. Curiously, the recently discovered ultra-faint satellite
Segue 2 \citep{Belokurov2009} appears immersed in the debris of what
very well might be the Tri-And cloud.

The recently discovered Cetus Polar Stream \citep{Newberg2009} has
avoided detection thanks to its low density and the overlap in
projection with much brighter Sagittarius trailing stream. However,
taking advantage of the SDSS spectroscopy available over a large
portion of the Southern Galactic sky, \citet{Newberg2009} present a
convincing argument in favor of a distinct stellar sub-structure,
colder and more metal-poor than the Sgr debris. \citet{Koposov2012}
provide the first sky map of the Cetus Polar Stream debris, and having
obtained accurate measurements of the stream's distance and velocity
gradients they argue that the sense of direction of the orbital motion
of the CPS is opposite to that of Sgr. In their maps, the structure
appears to be at least 20$^{\circ}$ wide and some 40$^{\circ}$ long,
yet with only 0.1 mag width along the line of sight.

The charting of the Galactic halo at distances beyond 50 kpc has been
somewhat sluggish due to the obvious lack of suitably bright tracers
covering a large enough area of the sky. A small fraction of the SDSS
footprint, so-called Stripe 82 has been imaged repeatedly during the
Supernovae campaign. \citet{Watkins2009} explores this multi-epoch
dataset to identify RR Lyra stars. They find a significant
over-density of RR Lyrae in the constellation of Pisces at
galacto-centric distances of $D\sim 90$~kpc, thus discovering the most
distant sub-structure known in the Milky Way halo. \citet{Sesar2010a}
confirm the discovery with a more sophisticated analysis of the same
SDSS data, while \citet{Kollmeier2009,Sesar2010b} present the
spectroscopic confirmation of the structure by obtaining velocities
for several RR Lyra members. As of today the true extent of the Pisces
Over-density is not known, but from the distribution of the RR Lyrae
it subtends at least $10^{\circ}$ on the sky making it some 15 kpc
wide.

\subsection{Quantifying the amount of sub-structure}

Within the $\Lambda$CDM paradigm, the global properties of the
Galactic stellar halo, namely the total luminosity, the shape, the
radial profile as well as the amount of sub-structure are simply the
consequences of the Milky Way's accretion history and as such all have
a straightforward interpretation. Observationally, however, these
properties are awkward to pinpoint. For example, to gauge the
flattening and the shape of the radial density profile, data across
large portions of the Northern and the Southern Galactic sky are
required. With pencil-beam surveys, the halo flattening or, more
generally any deviation from spherical symmetry (e.g. triaxliaity), is
impossible to determine and there is always a good chance of hitting
unknowingly a stellar stream or a cloud, hence biasing the estimates
of the density profile. Yet, in photometric studies, a robust global
density model is vital when quantifying the amount of
sub-structure. As the density distribution in the 6D phase-space,
where the individual accreted fragments are readily identifiable, is
collapsed onto the 3 spatial dimensions (or sometimes 2.5 or 2), the
signal is diluted as a result of super-position of many
structures. Therefore, even a small bias in the background properties
can affect dramatically the amplitude of sub-structure. Of course, the
``background'' itself, in this picture, is nothing else but the
stellar debris jumbled up more efficiently. Accordingly, the global
law parameterizing the behavior of the background provides crucial
information in which the mass of the satellites contributing to it and
the time of their accretion is encoded.

\subsubsection{Spatial inhomogeneities}

With plenty of deep multi-band photometry in both Galactic
hemispheres, the SDSS is an ideal resource to use to infer the global
properties of such an immense structure as the Milky Way's stellar
halo. A series of fits to the principal Galactic components as traced
by the MS stars in the SDSS DR5 is presented in
\citet{Juric2008}. This sample is dominated by the faint MS dwarfs
and, therefore, can not trace the volume density in the Milky Way much
further than 20 kpc. Within this radius, the halo appears to be well
described by a single power law density model with the index $n\sim
2.8$. Importantly, this study confirms earlier indications of a
substantial vertical flattening of the stellar halo $q\sim0.6$. The
results of \citet{Juric2008} are corroborated by the modelling of the
SDSS DR8 data with increased Southern Galactic hemisphere coverage
published by \citet{Bonaca2012b}. An attempt to delve deeper into the
stellar halo can be found in \citet{Bell2008}, where a simple
color-cut (similar to that illustrated in the right panel of
Figure~\ref{fig:bhb_msto}) is used to isolate the brightest of the old
MS stars in the halo. Using these blue, metal-poor turn-off stars,
with typical $M_g \sim 4$, it is possible to discern halo structures
as far as 30-40 kpc away from the Sun. However, as explained in
Section~\ref{sec:abs_mag}, the spread in the intrinsic luminosities of
the stars selected is as large as 3 magnitudes. There are two
important consequences of such blurred vision. First, convolving the
stellar halo distribution with large non-Gaussian errors in tracer
distances can have strong destructive effects on the accuracy of the
volume density inference. Second, when estimating the amplitude of
small scale deviations from the background, a debris at one particular
distance appears in several apparent magnitude bins (and hence
distances), thus biasing high the total amount of sub-structure across
the range probed. This effect is exacerbated at magnitudes close to
the survey limit, as well as for stars with different age and/or
metallicity.

While troubled by a number of issues outlined above, the analysis of
\citet{Bell2008} is the first of its kind. Taking advantage of the
impressive sky coverage and depth of the SDSS imaging, they provide a
quantitative interpretation of the inhomogeneous stellar halo glimpsed
by the earlier works. The main conclusions of the study by
\citet{Bell2008} are as follows. First, a smooth density model for the
MSTO tracers within 40 kpc is not appropriate for the Milky Way halo,
with most of the model parameters poorly constrained (see their
Figures 4, 7 and 9). Second, even after excising the major known
debris pile-ups such as Sagittarius stream and Virgo overdensity, the
amount of sub-structure $\sigma/{\rm total}$, parameterized in terms
of the scaled rms deviation $\sigma$ of the data around the smooth
model, stays just under $40\%$ from $r\sim 19$ mag to $r \sim 22$
mag. In the presence of these large stellar halo structures, the
$\sigma/{\rm total}$ statistic grows with apparent magnitude (roughly
proportional to distance) and reaches $>50\%$ at $r\sim
21.5$. Finally, \citet{Bell2008} compare the values of $\sigma/{\rm
  total}$ for the Milky Way halo traced by faint metal-poor MSTO stars
in the SDSS to those obtained for the semi-analytic stellar halo
simulations of the Galaxy by \citet{Bullock2005}. The 11 model halos
are made entirely of accreted stars, and show a minimal level of
sub-structure $\sigma/{\rm total} > 20\%$. Accordingly, the final
verdict is: the amount of sub-structure in the Galactic halo matches
that in the hierarchical galaxy formation models, and, therefore,
satellite accretion is the primary mode of the Milky Way's halo
creation.

\citet{Helmi2011} aims to improve the analysis of \citet{Bell2008} by
i) coming up with a more robust sub-structure quantification, and ii)
comparing the SDSS data to the most recent stellar halo
simulations. Similarly to \citet{Bell2008} they measure the stellar
density scatter in bins of apparent magnitude and the two celestial
coordinates. However, instead of calculating the amount of residual
deviation between the data and the best-fit smooth parametric model,
\citet{Helmi2011} work out the RMS around the mean stellar density in
the bin. Predictably, the amount of sub-structure computed in this
fashion is lower compared to that obtained by \citet{Bell2008}, albeit
only slightly. According to \citet{Helmi2011}, across the apparent
magnitude range of $18.5 < r < 22.5$, the normalized scatter ${\rm
  rms}(\rho)/<\rho>$ in the SDSS DR7 MSTO star density is at the level
of 30\% to 40\%. These rather serious levels of inhomogeneity found in
the SDSS data nonetheless appear low when contrasted with the degree
of sub-structure in simulated stellar halos. For the comparison with
the data, \citet{Helmi2011} examine the smoothness of the mock stellar
halos produced by \citet{Cooper2010}. These are built into the
Aquarius DM-only halos \citep{Springel2008} by tagging 1\% of the
most-bound particles in selected sub-halos and following them to
redshift 0. Compared to the mock ``Milky Ways'' of
\citet{Bullock2005}, these have the obvious advantages of being
fabricated in the Cosmological setting, and with a superior
resolution. However, there are disadvantages too. First, the Aquarius
suite explores only half as many accretion histories, in fact, in the
end, there are only 4 stellar halos analyzed in \citet{Helmi2011},
compared to 11 in \citet{Bullock2005}. Second, these Galaxy analogs do
not posses disks. A quick glance at the Figure 5 of \citet{Helmi2011}
reveals: all stellar halos of \citet{Cooper2010} are highly
irregular, with $50\% < \frac{{\rm rms}(\rho)}{<\rho>} < 150\%$. The
authors raise concern that some of the data-model discrepancy could be
due to the combined effects of the MSTO sample contamination and the
presence of a smooth, in-situ formed stellar halo
component. Nonetheless, they conclude that there exists considerable
tension between the observations of the Galactic stellar halo
sub-structure and the predictions of the simple but high-resolution
model. Even though the halos of both the real and the mock Galaxy are
very inhomogeneous, the simulations easily reach 2-3 times the
observed scatter on scales as small as few degrees.

\begin{figure}
\centering
\includegraphics[width=1.02\linewidth]{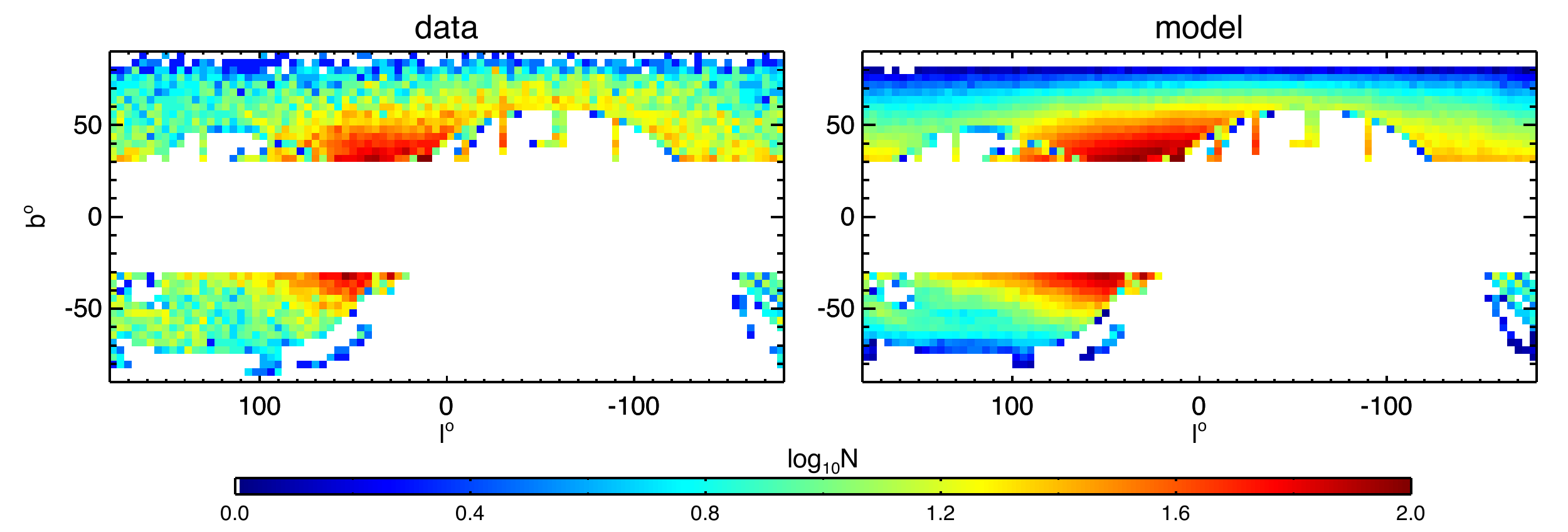}
\caption{Stellar halo of the Milky Way traced by the BHB stars. {\it
    Left} Distribution of the SDSS DR8 BHB candidates in the Galactic
  $l$ and $b$. {\it Right} Best-fit model of the stellar halo density
  distribution shown in the Left panel, from \citet{Deason2011a}. The
  model halo is flattened with $q\sim 0.6$ and has a break in the
  radial density profile at $r\sim 27$ kpc where the power-law index
  changes from -2.3 to -4.6. Figure courtesy of Alis Deason,
  IoA/UCSC.} \label{fig:halo_bhb}
\end{figure}

The picture of the utter chaos in the inner parts of the Galactic
stellar halo is re-visited in \citet{Deason2011a}. Instead of using
the more abundant MSTO stars, they choose to model the halo volume
density with Blue Horizontal Branch stars. While these stars are
rarer, their higher intrinsic luminosities, lower levels of
contamination and accurate absolute magnitude calibration independent
of age and chemistry all make these a better fit for the task. There
are, nonetheless, several limitations to the use of BHBs as
tracers. For example, being some $\sim 4$ magnitudes brighter and at
least two orders of magnitude less frequent as MSTO, these come in
particularly low numbers at bright apparent magnitudes due to the size
of the volume probed. Additionally, while their blue color makes them
stand out dramatically compared to most other stellar populations at
high Galactic latitudes, there is one troublesome impostor. Blue
Stragglers (see Figure~\ref{fig:tracers}) have close to identical
broad-band colors but are $\sim 1.5$ mag fainter. Outnumbering the
BHBs by a factor of 2 on average, these may pose a serious problem by
scrambling the tracer counts as a function of apparent
magnitude. \citet{Deason2011a} solve both the problem of the limited
dynamic range and of the contamination by including the BS stars in
the model. For all ``blue'' stars in the SDSS DR8, i.e. $-0.25 < g-r <
-0$, the probability of belonging to the BHB or the BS population is
assigned based on their $u-g$ and $g-r$ colors. As a result, the
number density of stars in volume elements of the space spanned by
position of the sky, color and apparent magnitude can be modeled
simply as the sum of the contributions from BHBs and BSs, weighted by
their conditional probabilities.

The results of the maximum-likelihood analysis presented in
\citet{Deason2011a} are summarized for the impatient reader in the
article's title ``Squashed, broken but smooth''. In other words: out
to 40 kpc, the Galactic stellar halo appears to be highly flattened,
the density profile follows closely the broken power law and, most
interestingly, the overall level of sub-structure detected using the
BHB tracers is rather low. At small and intermediate distances,
$\sigma/{\rm total}$ rises from as low $10\%$ to at most $20\%$
irrespective of the spatial scale of density perturbations. At large
distance, $\sigma/{\rm total}$ is close to $20\%$ on most scales, but
rises to $40\%$ for the angular sizes of several hundreds of
degrees. These numbers are obtained by excluding from the modeling the
regions of the sky with known large-scale halo overdensities. Even
when these are included, the small-size inhomogeneities are only $10
\% < \sigma/{\rm total} < 30\%$. While, superficially, these estimates
differ significantly from those quoted in \citet{Bell2008}, there are
several possible solutions to this discrepancy. Both methods have
their weak points. It is quite likely that some of the halo mess
observed by \citet{Bell2008} is simply due to the limitations of the
MSTO stars as tracers. On the other hand, the average number of BHBs
in a $1^{\circ} \times 1^{\circ}$ pixel is small, hence limiting the
areas of the sky tested by \citet{Deason2011a} to those towards the
inner Galaxy where mixing is more efficient. On slightly larger
angular scales (several degrees or so), it is, however, safe to
conclude that the inner stellar halo is indeed smooth.

\subsubsection{Phase-space sub-structure. Spaghetti, ECHOS and SKOs}

As it is much easier to identify the accreted satellite debris in the
phase-space compared to simple sky density maps or 3D spatial maps,
several attempts have been made to search for the surviving Galactic
sub-structure in the datasets of wide area spectroscopic surveys. The
Spaghetti survey \citep[e.g.][]{Morrison2000} is the first brave
endeavor to collect substantial numbers of genuine halo tracers in a
large distance range. It is set up to gather photometry and the
follow-up spectroscopy in several tens of ``pencil-beam'' fields over
the area covering many tens of degrees. The analysis dealing with the
quantification of the presence of sub-structure in the final set of
101 giants with spectra covering distances up to 100 kpc is presented
in \citet{Starkenburg2009}. They report the detection of 1 group and 6
pairs of clumped stars and conclude that their findings of 10\% of
sub-structure in the halo are consistent with the accretion scenarios
in which early and/or massive satellite infall leads to the creation
of broad phase-space features.

The SEGUE survey that has taken $\sim$240,000 spectra in $>$200
pointings spread over $\sim$11,000 square degrees is the ideal source
of data to carry out a systematic search for un-relaxed
sub-structure. \citet{Schlaufman2009} do exactly that, and detect in
137 lines of sight studied 10 high-confidence ECHOS, elements of cold
halo substructure as traced by metal-poor MSTO stars with distances in
the range $10 <D~({\rm kpc})< 20$. As the ECHOS identification
algorithm described is automated, the work also contains the results
of the completeness calculation for the sub-structure search in the
velocity space. These are then used to turn the numbers of detections
into the predictions of sub-structure fractions existent in the halo:
they conclude that within 1/3 of the volume of the Galactic halo,
there ought to be of the order of 10\% of ECHOS.

\citet{Xue2011} use the spectroscopic sample of $\sim 4,000$ SDSS BHB
stars to gauge the power spectrum of the kinematic sub-structure in
the stellar halo by measuring the excess of close stellar pairs whose
relative distances are measured in the 4D comprised of 3 spatial
coordinates and the line-of-sight velocity. The presence of
sub-structure is clearly detected when the observed close-pair
distribution is compared to that of the mock smooth halo created by
scrambling the heliocentric distances of the SDSS BHBs. Moreover, as
predicted, the outer halo ($20 < D < 40$ kpc) exhibits a stronger
sub-structure signal as compared to the inner one ($5 < D < 20$
kpc). However, when comparing to the simulations of
\citet{Bullock2005}, \citet{Xue2011} find appreciably less signal
overall, in particular, at the intermediate 4D distance scales.

Finally, at moderate distances from the Sun, relatively accurate
stellar proper motion measurements have been available for some
time. Combined with photometric parallax, this opens up the
possibility to calculate all 6 of the phase-space
coordinates. Accordingly, \citet{Smith2009} exploit the accurate
proper motions calculated by \citet{Bramich2008} using the multi-epoch
data in the SDSS Stripe 82 region, to compile a catalog of $\sim$
1,700 MS sub-dwarfs within 5 kpc from the Sun with full space
velocities and 3D coordinates measured. In estimating the tangential
components of the stars velocity, the main source of error is the
distance uncertainty, which is actually present at the rather modest
level of 10\%. After subtracting a smooth Galaxy model from the
distribution of the angular momenta $J_z, J_{\perp}$ 3 significant
clumps are detected, one of which has been previously discovered in
the pioneering work by \citet{Helmi1999}.

\section{Putting the puzzle together}

It is obvious why in a framework which builds the Universe from bottom
up, the smallest galaxies are under double scrutiny: such structure
formation paradigm risks losing its credibility if its basic blocks
look unlifelike. For $\Lambda$CDM, predicting the properties of the
local dwarf galaxies, while seemingly completely straightforward, has
turned out to be a trying quest. As this review argues, before a
faithful comparison between the dwarf galaxies formed according to the
Cold Dark Matter model and the observed satellites can be carried out, a
number of loose ends need to be tied up. First, the host-to-host
variations in the properties of the DM sub-halo populations due to the
parent halo accretion history and the environment need to be
quantified. Also, it is not an overstatement to say that our
understanding of the baryonic processes to do with cooling and
feedback, although developing swiftly, is not yet mature and, thus
deserves further improvement. Observationally, it seems now possible
to provide better measurements of the basic properties of the host
such as the Galactic total mass, the mass of the disk and, possibly
even the concentration of the DM halo. Additionally, as the
quantitative comparison is presently based around the satellite
luminosity function, it is crucial to figure out the genesis of the
smallest objects dominating the census, the ultra-faints. Finally, it
is clear that the same DM and baryonic physical processes that have
sculpted the $z=0$ dwarf satellite population are also responsible for
the creation of the stellar halo. Therefore, a successful galaxy
formation model is expected to get both the survived and the perished
satellites right.

\subsection{Light Galactic DM halo with high concentration} 

\citet{Deason2012b} tackle the issue of the dearth of reliable tracers
in the outer halo.  They take advantage of the availability of the
multiple epoch imaging data across the SDSS field of view. Due to the
survey geometry, a considerable number of imaging ``stripes'' overlap
(mostly around the survey poles) and the photometry from the multiple
runs can be combined to yield higher signal-to-noise magnitude
measurements.  This is especially valuable for the faint BHB candidate
stars, whose single-epoch $u$-band magnitudes are simply too
unreliable for the conventional broad-band BHB/MS/MSTO
classification. Using the overlaps as well as the multi-epoch data
from the SDSS Stripe 82, \citet{Deason2012b} find 43 distant BHB
candidates with $20 < g < 22$ and follow these up with deep
spectroscopy on the ESO's VLT. Once a handful the QSO interlopers are
removed, of the remaining 38 A-colored stars, 7 are genuine BHBs in
the distance range 80-150 kpc. Note that, at these faint magnitudes,
even the remaining 31 BS stars populate the poorly explored regime
between 30 and 90 kpc. Their final sample also contains 8 cool carbon
stars that span a distance range 80 to 160 kpc. This combined
spectroscopic sample is the largest collection to date of the halo
tracers with distances beyond 60 kpc. Curiously, outside 100 kpc from
the Galactic center, the line-of-sight velocity dispersion plummets to
a rather low 50-60 km s$^{-1}$. Unless the stellar tracers considered
have a significant tangential bias and/or their density drops much
faster than the already rather steep power law with index -4.5, this
measurement implies the mass of the Milky Way in the range of $5-10
\times 10^{11} M_{\odot}$ within 150 kpc.

To figure out how concentrated the DM halo of the Galaxy is, the total
matter density has to be mapped out in a sufficiently wide range of
distances around or beyond the Solar neighborhood (inside $R_0$, the
disk dominates the mass budget). \citet{Deason2012a} exploit the sheer
volume probed by the SDSS BHB stars with known radial velocities to
simultaneously constrain the halo velocity anistropy and the matter
distribution. By approximating the gravitational potential of the
Galaxy as a power-law and adopting the tracer density distribution
pinned down in \citet{Deason2011a}, they find that the rotation curve
of the Galaxy has to start falling appreciably already at around 30
kpc from the center. They ague that the preferred value of the
normalization and the power law index of the gravitational potential
are inconsistent with massive, i.e. $2 \times 10^{12} M_{\odot}$ DM
halos with concentrations around $c\sim 10$. Instead, a lighter and
more concentrated, i.e. $c\sim 20$ dark halo is favored.

\subsection{Smooth stellar halo and signatures of early accretion}

\begin{figure}
\centering
\includegraphics[width=0.85\linewidth]{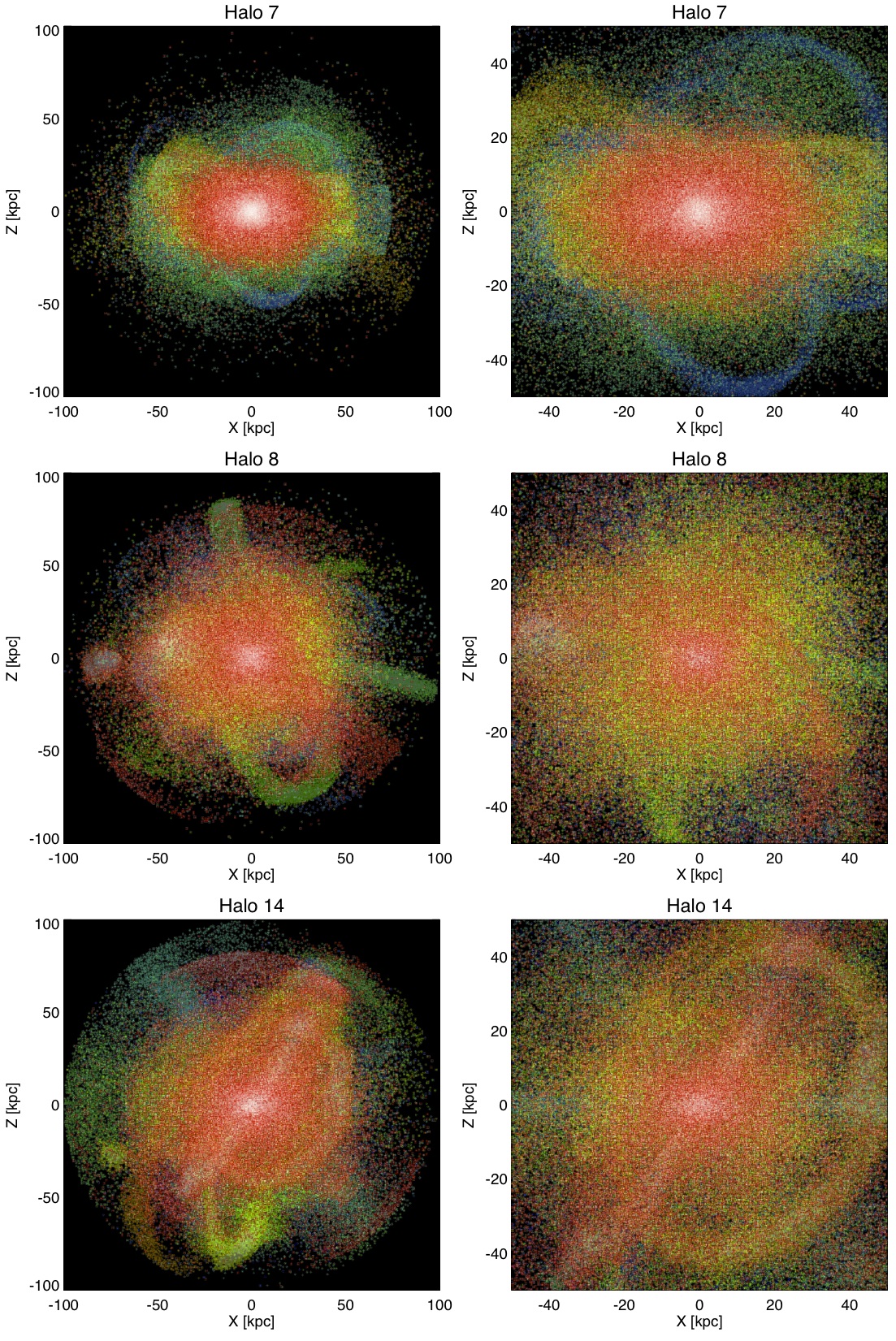}
\caption{Three of the 11 Galactic stellar halos simulated by
  \citet{Bullock2005}. Particles are color-coded according to the mass
  of the progenitor, with most massive in red and least massive in
  blue. The smooth, flattened Halo 7 has a density break at 24 kpc and
  represents the simulation closest to the observed Galactic stellar
  halo. Halo 8 gives a clue of how the stellar halo in M31 might look
  like.}

\label{fig:bj2005}
\end{figure}

The stellar masses in the known large-scale over-densities in the
Galactic halo (see Section~\ref{sec_big4}) sum up to the approximate
total of $\sim 3 \times 10^8 M_{\odot}$. This is to be compared with
the estimates of the total stellar halo mass \citep[e.g.][]{Bell2008,
  Deason2011a} of the order of $\sim 10 \times 10^8
M_{\odot}$. Therefore, perhaps as much as $70 \%$ of the stellar halo
within 40 kpc is in a smooth or, more accurately, well-mixed,
component, which can be described by a broken power-law density
profile with a flattening of $q\sim 0.6$. On contrary, the Andromeda's
stellar halo harbors no such feature in its density profile over twice
as large range of distances, albeit it does not look nearly as smooth
\citep{Gilbert2012}. \citet{Deason2013} investigate whether stellar
halos with density breaks similar to that of the Milky Way can be
assembled purely through satellite accretion, and if yes, what
controls the break prominence and the radius. By studying a set of 11
N-body simulations published by \citet{Bullock2005} they come to the
following conclusion. In the simulations studied, the radial profiles
of stars from individual accretion events can be described by single
power law, double power law or can be so ill-defined that neither of
the simple models works. Ancient ($>10$ Gyr \footnote{In this
  definition, the age marks the time of when the stars became unbound,
  which implies slightly earlier epochs for the arrival of the
  progenitor.}) debris have had plenty of time to mix and therefore at
$z=0$ the radial profile is comfortably fit with a single
power-law. Old (7-10 Gyr) debris have spread out over a range of
Galacto-centric distances but around the progenitor's apo-centre, the
drop in stellar density remains. Recent ($<6$ Gyr) mergers have not yet
filled the entire volume inward of the apo-centre and their radial
distribution still peaks at $R>0$.

The stellar halo (in this model) is just a superposition of the debris
from the individual events across the entire accretion history. The
combined stellar profile can have a distinct break (at the average
apo-centre of the most massive accreted satellites) only if the most
significant merger(s) happened at the right time, i.e. 8-10
Gyr. Additionally, it is required to dampen the accretion rate at the
subsequent epochs: as the Galaxy grows, the satellites that arrive
with increasingly larger apo-centers thus can flatten out the density
profile around and beyond the break radius, thus erasing this feature
altogether. The hypothesis that the density break in the Galactic
stellar halo reflects the apo-center(s) of the massive satellite(s)
accreted at early epochs can be tested with 3D kinematics. Radial
velocities of stars tend to zero around the apo-center of the orbit,
therefore the radial velocity dispersion of the stellar halo should
have a dip around the break radius as well as an increase in the
tangential anisotropy. Moreover, there exists a potentially powerful
diagnostic to decipher the properties of this old merger. Namely, if
the metallicities of the stellar halo tracers around the break radius
(i.e. $20<R<30$ kpc) are available, then it is possible to distinguish
between the accretion of one or two massive satellite(s) and the
accretion of a group of dwarfs. If only one system contributed the
bulk of the debris within the break then the radial velocity
dispersion dip around the break radius should be most visible in stars
with the chemical abundance of that satellite. The density break
created via superposition of the debris from many different satellites
is not dominated by any particular stellar population, and hence, the
drop in radial velocity dispersion should occur, albeit weakly, across
the metallicity range.

Figure~\ref{fig:bj2005} shows the examples of three Galactic stellar
halos created in simulations by \citet{Bullock2005}. Here, the X-Z
distributions of particles color-coded according to the mass of the
progenitor (red for most massive, blue for the least massive) are
shown for the inner 100 kpc (left panel) and the inner 50 kpc (right
panel). The differences in the structures of Halo 7 and Halo 8 provide
visual clues as to the findings of \citet{Deason2013}. Halo 7 has a
peaked accretion history, with the bulk of the stellar halo assembled
quickly around 8 Gyr ago. Within 40 kpc from the center of the Galaxy,
the stellar halo seems smooth, flattened and strongly aligned with the
disk (compare to the view of the Galactic stellar halo in
Figure~\ref{fig:halo_bhb}). Note that Halo 7 has a prominent break in
the radial density profile at 24 kpc, closely matching the observed
properties of the Milky Way halo as traced by the BHBs. Halo 8, on
contrary, shows no evidence of the break - this turns out to be a
giveaway of its accretion history and the present day appearance. Halo
8 can be categorized by a continuous infall of satellites with three
particularly massive fragments disrupting at 11, 7 and 1-2 Gyr. The
stellar distribution is evidently more extended and substantially
messier (even in the inner parts) as compared to that of the ``quiet''
Halo 7. Additionally, neither obvious vertical flattening nor
alignment with the disk can be seen.

\subsection{Nature of the faintest of the ultra-faint satellites}

\begin{figure}
\centering
\includegraphics[width=0.99\linewidth]{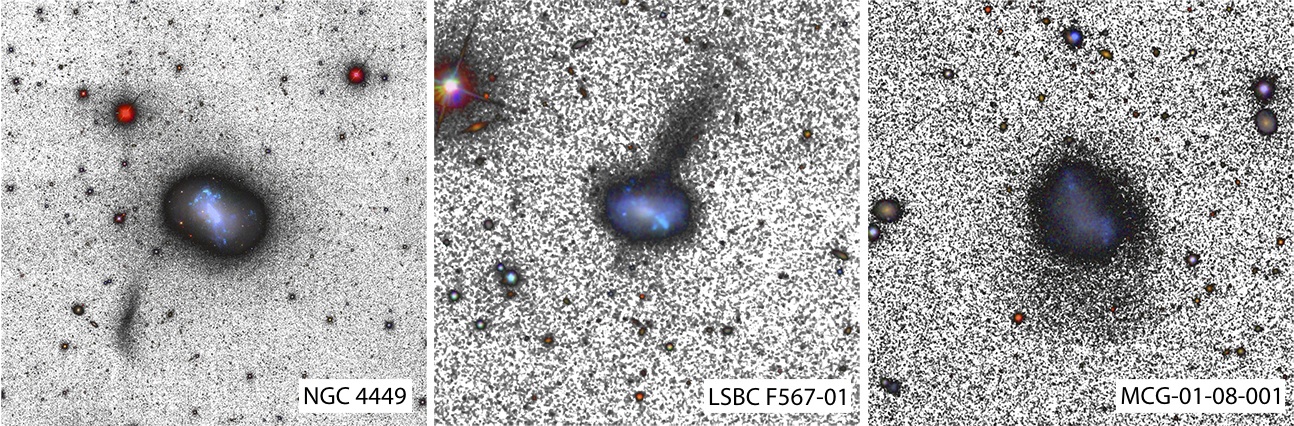}
\caption{Dwarfs accreting dwarfs. This shows the SDSS image mosaics of
  the three Local Volume dwarf galaxies (with luminosities similar to
  that of the SMC) disrupting their satellites with masses as low as a
  hundredth of the host's. The images are processed to enhance the
  low-surface brightness features. From Left to Right: the
  dwarf-to-dwarf accretion sequence. The satellite of the NGC 4449 is
  showing the first signs of tidal interaction, the satellite of LSBC
  F567-01 is critically stretched, very close to being fully
  destroyed, the satellite of MCG-01-08-001 is taken apart and
  incorporated into the host, but still visible as faint stellar plume
  around the Southern edge of host's halo.}

\label{fig:dwarf_accretion}
\end{figure}

The absolute majority of the stellar populations in the ultra-faint
dwarf satellites are as old as the oldest Galactic globular clusters
and are similarly metal-poor, as revealed, for example, by their deep
Color-Magnitude Diagrams. Given their low total stellar masses, the
velocity dispersions in the range of $2 < \sigma < 9$ km s$^{-1}$ and
the recently detected spreads in metallicity, there is now little
doubt that these are the remnants of the galaxies born at high
redshifts in low-mass DM halos. What is not yet completely clear is
the exact mapping between the galaxies of a particular luminosity at
$z=0$ and the original sub-halo mass at the epoch of formation. This
is simply due to the fact that for such faint stellar systems, there
is not enough stars currently available to trace the total matter
distribution out to large distances from the center. Hence, even
though their present-day central masses are being constrained, the
total mass and the extent of their DM halos is still an enigma.

All semi-analytic models of the dwarf galaxy formation predict that
most satellites lose significant amounts of their dark and luminous
matter to the host's tides. The precise mass loss is the crucial
middle part in the model that has at one end a variety of dark hosts
with the mass function steeply rising at low masses, and at the other,
the observed population of Galactic survivors. Relying on the fact
that the details of the tidal harassment are captured at the
appropriate level in the high resolution N-body simulations, the
luminosity function of the Milky Way satellites is inverted to reveal
the physics of star formation in the early Universe. However, it is
now apparent that the inclusion of the baryons (perhaps even as simple
as adding a disk component to the host galaxy) can alter the satellite
survival rates dramatically. Additionally, in the hierarchical
Universe, the smallest satellites have a non-zero chance of being
first accreted onto the more massive dwarf galaxies before finally
merging with their ``Milky Way'' host. Such {\it satellites of
  satellites} with large host-to-satellite mass ratios have been known
to exist in the galaxy formation simulations, but only now begin to be
discovered in nature. For example, \cite{Martinez-Delgado2012} report
the discovery of the early stages of the accretion of a low-luminosity
dwarf by the galaxy as massive as the SMC, namely NGC
4449. Figure~\ref{fig:dwarf_accretion} shows this and two more
examples of different stages of the accretion of dwarfs onto dwarfs.

To increase the dynamic range, the images in the three panels of
Figure~\ref{fig:dwarf_accretion} are composed of two versions of the
same SDSS mosaic image: the original one in color and the greyscale
one obtained by applying the median filter to triples of pixels in the
three SDSS filters to minimize the ``patchwork'' effects of the sky
slightly misaligned between different SDSS runs. The original SDSS
color image and the stretched and inverted greyscale image are then
combined by choosing the pixel value to be the highest one between the
two images. This simple image processing \citep[also employed by
  e.g.][]{Martinez-Delgado2010, Martinez-Delgado2012} makes it
possible to show simultaneously the central object where most of the
light is as well as the further low surface brightness features that
are to do with the ongoing tidal interactions. While the left panel
shows the beginning of the process of accretion, the middle and the
right panels of the Figure display more advanced stages of the dwarf
infall and disruption. LSBC F567-01 in the center is seen taking apart
a low-luminosity satellite which looks prominently stretched and
perhaps close to its total disintegration. Finally, MCG-01-08-001 in
the right panel, is clearly about to finish eating up a smaller
satellite, which presently can only be seen as extended faint plume of
stellar debris in the Southern parts of the host.

Taking into account the existing evidence of a likely association
between objects like Segue 1, Segue 2, Bootes II and III and the known
large-scale sub-structure in the Galactic stellar halo, the following
scenario is perhaps possible. The faintest of the currently known
Milky Way satellites were born as sub-systems of larger dwarfs and
were subsequently pre-processed by their first hosts' tidal
fields. They have probably lost much of their original mass but their
remnant cores survived as part of the bigger dwarf galaxy until the
whole system was accreted by the Milky Way. Their parent galaxies were
sufficiently massive to be dragged into the inner Milky Way where they
were destroyed and are visible today only as streams and clouds of
stars in the halo. Some of the satellites of satellites persisted in
the Milky Way halo and can now be observed as the faintest of the
ultra-faint dwarfs. In this scheme, the current stellar and the DM
masses are severely truncated: thanks to the pre-processing in the
gravitational field of the parent dwarf their current luminosities
could be much lower than what is attainable by the lowest mass dwarfs
in the field. This in turn, would imply that the fast drop in the
efficiency of the dwarf galaxy formation actually happens at the
sub-halo masses that are higher than previously envisaged. Because the
satellites with absolute magnitudes around $-2 > M_V > -4$ can only
(or mostly) exist as part of bigger dwarf systems, their distribution
in the Galaxy is different from that of the accreted field dwarf
population. Their radial density profile should be strongly radially
concentrated due to the combination of the two effects. First, their
parent galaxies were massive enough to end up close to the center due
to the dynamical friction. Second, in the Galaxy most of the large
systems (apart from the Sgr dwarf) were accreted as early as 8-10 Gyr,
when the mass and the virial mass of the Milky Way were much
smaller. Taking these effects into account, much lower numbers of
satellites as faint as Segue I or II are predicted to be discovered by
the future deep all sky surveys.

\pagebreak

\section*{Acknowledgments} 
V. Belokurov thanks The Royal Society the support. The work on this
review has received funding from the European Research Council under
the European Union's Seventh Framework Programme (FP/2007-2013) / ERC
Grant Agreement n. 308024. The author has enjoyed conversations with
A. Deason, W. Evans, A. Helmi, M. Irwin, S. Koposov, P. Kroupa,
J. Norris, M. Smith, E. Starkenburg and E. Tolstoy.

\pagebreak

\vspace*{2cm}

\noindent

\bibliographystyle{elsarticle-harv}
\bibliography{belokurov_narev}

\end{document}